\documentclass[11pt]{article}
\usepackage{a4}
\usepackage{amssymb}
\usepackage{cite}
\usepackage{epsf}\usepackage{amssymb} \usepackage{cite}
\usepackage{amsthm}
\usepackage{amscd}
\usepackage{amsfonts}
\usepackage{amsmath}
\usepackage{hyperref}
\newcommand{\be}{\begin{equation}}
\newcommand{\ee}{\end{equation}}
\newcommand{\bea}{\begin{eqnarray}}
\newcommand{\eea}{\end{eqnarray}}

\def\stackreb#1#2{\ \mathrel{\mathop{#1}\limits_{#2}}}

\begin{document}
\thispagestyle{empty}
\def\thefootnote{\fnsymbol{footnote}}
\begin{center}\Large
S-move matrix in the NS sector of \\ $N=1$ super Liouville field theory\\
\end{center}

\vskip 0.2cm

\begin{center}
Elena Apresyan$^{1,2}$\footnote{elena-apresyan@mail.ru}
and
Gor Sarkissian$^{1,2}$
\footnote{sarkissn@theor.jinr.ru, gor.sargsyan@yerphi.am}
\end{center}
\begin{center}
$^1$Yerevan Physics Institute,\\
Alikhanian Br. 2, 0036\, Yerevan, Armenia
\end{center}
\begin{center}
and
\end{center}
\begin{center}
$^2$ Bogoliubov Laboratory of Theoretical Physics, JINR,\\
Dubna, Moscow region, 141980 Russia\\
\end{center}

\vskip 1.5em

\begin{abstract} \noindent
In this paper we calculate matrix of modular transformations of the one-point toric conformal blocks in
the Neveu-Schwarz sector of $N=1$ super Liouville field theory. For this purpose we use explicit expression for this
matrix as integral of product of certain elements of fusion matrix. This integral is computed using the chain of integral
identities for supersymmetric hyperbolic gamma functions derived by the degeneration of the integrals of parafermionic
elliptic gamma functions.
\end{abstract}

\newpage
\tableofcontents
\newpage

\section{Introduction}
One of the basic objects in two-dimensional conformal field theory is the $S$-matrix of modular transformation of one-point toric conformal blocks.

One can write  the one-point correlation function
on a torus with modular parameter $\tau$
in the following form:
 \be\label{onep}
 \langle \phi_{\lambda,\bar{\lambda}} \rangle = \sum_{(\gamma,\bar\gamma)} \mathcal{F}_{\gamma}^{\lambda}(q)\,
  \bar{\mathcal{F}}_{\bar{\gamma}}^{\bar{\lambda}}(\bar q) \, C^{\lambda,\bar{\lambda}}_{\gamma,\bar{\gamma}}\ ,
 \ee
where $q = e^{2 \pi i \tau}$, $\mathcal{F}_{\gamma}^{\lambda}(q)$ are one-point conformal toric blocks and $C^{\lambda,\bar{\lambda}}_{\gamma,\bar{\gamma}}$ are structure constants.

The $S$-matrix of modular transformation of the one-point toric blocks on a torus is defined by relation
\begin{equation}
\label{mmod:def}
{\cal F}_{\beta_1}^{\beta_3}(q(\tau))
 =
(-i\tau)^{-\Delta_{\beta_3}}
\sum_{\beta_2}  \;
S_{\beta_1\beta_2}(\beta_3)\,
{\cal F}_{\beta_2}^{\beta_3}\left(q\left(\textstyle -{1\over \tau}\right)\right)\, .
\end{equation}

The $S$-matrix is one of the most important notions of $2D$ conformal field theories. The $S$-matrix, together with fusion $F$-matrix, braiding $B$-matrix and $T$-matrix, representing the second generator of the modular group $\tau\to \tau+1$,
form the Moore-Seiberg groupoid \cite{Moore:1988qv}. These generators satisfy a set of relations called Moore-Seiberg's equations \cite{Moore:1988qv}. One of them will be discussed below in detail. The Moore-Seiberg groupoid gives in a sense the full number of operations on the space of conformal block. More abstractly one can consider the space of conformal blocks as a representation of
Moore-Seiberg groupoid. For better understanding of the Moore-Seiberg's equations we recall that the space of conformal blocks
in rational conformal field theory also arises as Hilbert space of states of $3D$ topological field theory
\cite{Witten:1988hf,Elitzur:1989nr}. It was observed that the Moore-Seiberg's equations are equivalent to three-dimensional topological invariance \cite{Witten:1988hf,Moore:1989vd}. In fact, it was proved for rational theories that given a representation
of the Moore-Seiberg groupoid, namely a solution to the Moore-Seiberg's equations, one can reconstruct the $3D$ topological field theory
\cite{pascal}.

The Moore-Seiberg groupoid proved to be useful also for non-rational theories.

It was proposed in \cite{Verlinde:1989ua} that the space of conformal block of Liouville theory with its mapping class
group representation is isomorphic to the Hilbert space of states of quantized Teichmuller space. Later this proposal was refined
in \cite{Teschner:2003at,Teschner:2005bz} and it was shown
that quantized Teichmuller space carries unitary projective representation of Moore-Seiberg grupoid
equivalent to that of Liouville theory.
 Using these data  it was built in \cite{Collier:2023fwi,Collier:2024mgv}, in full analogy with rational case, topological field theory
 based on conformal blocks of the Liouville theory, called Virasoro topological quantum field theory. In these papers this Virasoro
 topological quantum field theory was related to $3D$ quantum gravity with negative cosmological constant.

The $S$-matrix $S_{\beta_1\beta_2}(\beta_3)$ satisfies the following Moore-Seiberg duality identity \cite{Moore:1988qv,Moore:1989vd}:
 \bea\label{msrel}
&&S_{\beta_1\beta_2}(\beta_3)\sum_{\beta_4}F_{\beta_3,\beta_4}\left[\begin{array}{cc}
\beta_2&\alpha_1\\
\beta_2&\alpha_2 \end{array}\right]e^{-2\pi i (\Delta_{\beta_4}-\Delta_{\beta_2})}
F_{\beta_4,\beta_5}\left[\begin{array}{cc}
\alpha_2&\alpha_1\\
\beta_2&\beta_2 \end{array}\right]\\ \nonumber
&&=\sum_{\beta_6}F_{\beta_3,\beta_6}\left[\begin{array}{cc}
\beta_1&\alpha_1\\
\beta_1&\alpha_2 \end{array}\right]e^{-\pi i (\Delta_{\alpha_1}+\Delta_{\alpha_2}-\Delta_{\beta_5})}
F_{\beta_1,\beta_5}\left[\begin{array}{cc}
\alpha_1&\alpha_2\\
\beta_6&\beta_6 \end{array}\right]S_{\beta_6\beta_2}(\beta_5)\, .
\eea
Here $F_{\sigma_5,\sigma_6}\left[\begin{array}{cc}
\sigma_2&\sigma_1\\
\sigma_3&\sigma_4 \end{array}\right]$ is fusion matrix and $\Delta_{\sigma}$ is conformal weight.
Setting in eq. \eqref{msrel} $\beta_1=\beta_3=0$  one obtains an
explicit expression of the $S$-matrix in terms of the fusion matrix and conformal weights  \cite{Behrend:1999bn,Moore:1989vd}:
\bea
&&S_{0\beta_2}\sum_{\beta_4}F_{0,\beta_4}\left[\begin{array}{cc}
\beta_2&\alpha\\
\beta_2&\alpha \end{array}\right]e^{-2\pi i \Delta_{\beta_4}}
F_{\beta_4,\beta_5}\left[\begin{array}{cc}
\alpha&\alpha\\
\beta_2&\beta_2 \end{array}\right]\\ \nonumber
&&=e^{-\pi i (2\Delta_{\alpha}-\Delta_{\beta_5}+2\Delta_{\beta_2})}
F_{0,\beta_5}\left[\begin{array}{cc}
\alpha&\alpha\\
\alpha&\alpha \end{array}\right]S_{\alpha\beta_2}(\beta_5)
\eea
or this can be written also in the form \cite{Behrend:1999bn}
\bea\label{behr}
&&S_{0\beta_2}\sum_{\beta_4}F_{0,\beta_4}\left[\begin{array}{cc}
\beta_2&\alpha\\
\beta_2&\alpha \end{array}\right]e^{2\pi i \Delta_{\beta_4}}
F_{\beta_4,\beta_5}\left[\begin{array}{cc}
\alpha&\alpha\\
\beta_2&\beta_2 \end{array}\right]\\ \nonumber
&&=e^{\pi i (2\Delta_{\alpha}+2\Delta_{\beta_2})}
F_{0,\beta_5}\left[\begin{array}{cc}
\alpha&\alpha\\
\alpha&\alpha \end{array}\right]S_{\alpha\beta_2}(\beta_5)\, .
\eea
Here we used that $S_{\alpha\beta_2}(\beta_5)$ in the limit $\beta_5\to 0$ becomes matrix $S_{\alpha\beta_2}$ of the modular transformation of characters and $S_{0\beta_2}$ in \eqref{behr} is the corresponding element of this matrix. The formula (\ref{behr}) was used in \cite{Vartanov:2013ima} to calculate $S_{\beta_6\beta_2}(\beta_5)$ in the Liouville field theory.
The calculations are very involved and use sequence of integral identities for various products of hyperbolic gamma functions.
Important point here is that all the necessary identities can be derived by some chains of the reductions from the
elliptic beta integral  \cite{spi:umn} and symmetry relations of so called elliptic $V$-function \cite{spi:theta,spi:essays}.
This becomes possible since as shown in \cite{ruij,rai:limits} the hyperbolic gamma function is an
asymptotic form of the elliptic gamma function.

As a result it was obtained that the $S^{L}_{\alpha\beta_2}(\beta_5)$-matrix in Liouville theory is given by the expression
\cite{Teschner:2003at,Vartanov:2013ima}:
\be
S^{L}_{\alpha\beta_2}(\beta_5)={{\cal N}(\alpha,\beta_5,\alpha)\over {\cal N}(\beta_2,\beta_5,\beta_2)}
S^{L}_{0\beta_2}e^{-{i\pi\over 2}\Delta^L_{\beta_5}}S^{T}_{\alpha\beta_2}(\beta_5)\, ,
\ee
where ${\cal N}(\alpha_1,\alpha_2,\alpha_3)$ are some normalization coefficients of chiral vertex operators, 
$\Delta^L_{\alpha}=\alpha(Q-\alpha)$ is conformal weight, $Q=b+1/b$,
\be
S_{0\alpha}^L={1\over \sqrt{2}S_b(\pm 2(\alpha-Q/2))}=-2\sqrt{2}\sin{\pi (2\alpha-Q)\over b}\sin\pi (2\alpha-Q)b
\ee
is the corresponding component of the matrix of modular transformation of characters, and
\bea\label{partfun}
&&S^{T}_{\alpha\beta_2}(\beta_5)={1\over S(\beta_5)}\int_{-i\infty}^{i\infty}S_b(y-Q/2+\beta_2+\beta_5/2)
S_b(y+Q/2-\beta_2+\beta_5/2)\qquad\\ \nonumber
&&\times S_b(-y-Q/2+\beta_2+\beta_5/2)
S_b(-y+Q/2-\beta_2+\beta_5/2)e^{2i\pi y(2\alpha-Q)}  dy\, .
\eea
Here $S_b(x)$ is hyperbolic gamma function defined in appendix A. For the physical states all the entering
parameters belong to the set $Q/2+i\mathbb{R}$.

The function \eqref{partfun} gives the bulk-boundary
coupling in LFT as well \cite{Hosomichi:2001xc}, since the sewing constraints imply that the bulk-boundary coupling proportional to
$S^{L}_{\alpha\beta_2}(\beta_5)$ \cite{Runkel:1998he,Behrend:1999bn}.

Using identities \eqref{tigr} and \eqref{symref} in appendices B and E correspondingly for $r=1$ one can show
that this function enjoys the following properties
\be\label{pent1}
S^{T}_{\alpha\beta_2}(\beta_5)=S^{T}_{\beta_2\alpha}(Q-\beta_5)
\ee
and
\bea\label{pent2}
&&S^{T}_{\alpha\beta_2}(Q-\beta_5)=S_b(-Q+2\alpha+\beta_5)S_b(Q-2\alpha+\beta_5)\\ \nonumber
&&\times S_b(2Q-\beta_5-2\beta_2)S_b(-\beta_5+2\beta_2)
S^{T}_{\alpha\beta_2}(\beta_5)\, .
\eea

Note that function \eqref{partfun} is nothing else as partition function of mass-deformed $N=4$ SQED with two electron hypermultiplets,
called $T[SU(2)]$, on squashed three-sphere $S^3_b$ and eq.\eqref{pent1} expresses its mirror symmetry  \cite{Hama:2010av,Hosomichi:2010vh}.
This coincidence is not accidental but consequence of AGT dualty \cite{Alday:2009aq} between $4D$ $N=2$ $SU(2)$ supersymmetryc gauge theory
on $S^4$ and Liouville theory
in the presence of domain walls/defects \cite{Drukker:2010jp,Hosomichi:2010vh}. It was proposed in \cite{Drukker:2010jp} that partition  function of $3D$ S-duality domain wall inside $4D$ $N=2^*$
theory, which is the mentioned mass-deformed $T[SU(2)]$ theory, should coincide with the kernel of $S$-move transformation of Liouville theory.
The actual coincidence was established in  \cite{Hosomichi:2010vh}.

It is possible to show that the invariance of the one-point correlation function \eqref{onep} under the modular transformation
requires $S$-matrix to satisfy the following Moore-Seiberg equation (see for example \cite{Collier:2023fwi,Vartanov:2013ima}):
\be\label{msorto}
\int d\beta_2 S^L_{\beta_1\beta_2}(\beta_5)S^L_{\beta_2\beta_3}(\beta_5)=\delta(\beta_1-\beta_3)e^{-i\pi\Delta^L_{\beta_5}}\, .
\ee
It is easy to check that using the relations \eqref{pent1} and \eqref{pent2} the eq.\eqref{msorto} can be written as
\bea\label{msorto2}
&&\int d\beta_2 S^T_{\beta_2\beta_1}(\beta_5)S^T_{\beta_2\beta_3}(\beta_5)S^{L}_{0\beta_2}S_b(-Q+2\beta_2+\beta_5)S_b(Q-2\beta_2+\beta_5)
\\ \nonumber
&&={1\over S^{L}_{0\beta_1}}\delta(\beta_1-\beta_3)S_b(-Q+2\beta_1+\beta_5)S_b(Q-2\beta_1+\beta_5)\, .
\eea

This property is closely related to the fact that the function $S^T_{\alpha\beta_2}(\beta_5)$ satisfies the following eigenvalue equation \cite{Galakhov:2013jma,Nemkov:2015zha}
\be\label{ruijs}
H(S^T_{\alpha\beta_2}(\beta_5))=2\cos\pi b(Q-2\beta_2)S^T_{\alpha\beta_2}(\beta_5)\, ,
\ee
where $H$ is the finite-difference operator:
\bea\label{hamo}
H={\sin\pi b(2\alpha-Q-\beta_5)\over \sin \pi b(2\alpha-Q)}e^{-{b\over 2}\partial_{\alpha}}+
{\sin\pi b(2\alpha-Q+\beta_5)\over \sin \pi b (2\alpha-Q)}e^{{b\over 2}\partial_{\alpha}}\, .
\eea
This equation can be derived in the numerous ways. First one can obtain it evaluating the Moore-Seiberg equation \eqref{msrel} for
degenerate primaries \cite{Nemkov:2015zha}. It is possible directly to check \cite{Galakhov:2013jma} that $S^T_{\alpha\beta_2}(\beta_5)$ satisfies
\eqref{ruijs}. And finally one can derive it by certain limiting procedure from the difference equations discussed in \cite{Apresyan:2022erh}.
The operator \eqref{hamo} is well known Hamiltonian of the relativistic Calogero-Sutherland model \cite{ruijj}, sometimes also called Ruijsenaars-Schneider model, and was extensively studied recently is series of works \cite{Belousov:2023qgn,Belousov:2023pcy,Belousov:2023sat}.
Now presence of the $\delta$-function in the r.h.s. of \eqref{msorto2} is easily understood from the observation that the Hamiltonian \eqref{hamo} is hermitian w.r.t. the scalar product given by the l.h.s. of \eqref{msorto2}.
The integral in l.h.s. of  \eqref{msorto2} is computed in \cite{Belousov:2023qgn,Belousov:2023pcy,Belousov:2023sat}
and indeed coincides with r.h.s. of \eqref{msorto2}. We see that the function \eqref{partfun} is on the junction of three theories:
Liouviile field theory, $T[SU(2)]$ theory and relativistic Calogero-Sutherland model, and various its properties
have different interpretation in each of them.

In this paper we use the formula (\ref{behr}) to calculate $S$-matrix of modular transformation of one-point toric conformal block
in $N=1$ super Liouville field theory (SLFT).  For this purpose we should elaborate  ``supersymmetric counterpart" of the mentioned
 integral identities. At this point we would like to point out that if the essential part of the fusion matrix in the Liouville field theory is integral of product of  hyperbolic gamma functions $S_b(x)$, defined in appendix A, then the essential part of the fusion matrix in $N=1$ super Liouville field theory is sum of integrals
of product of so called  Neveu-Schwarz and Ramond hyperbolic gamma functions:
\be\label{nsf}
 S_{1}(y)\equiv S_{\rm NS}(y) \equiv S_b\left({y\over 2}\right)
S_b\left({y\over 2}+{Q\over 2}\right),
\ee
\be\label{rf}
 S_{0}(y)\equiv S_{\rm R}(y)\equiv S_b\left({y\over 2}+{b\over 2}\right)
S_b\left({y\over 2}+{b^{-1}\over 2}\right)\, ,
\ee
where $Q=b+1/b$.
To derive the necessary integral formulas for them first we note that the functions (\ref{nsf}) and (\ref{rf})
are particular case of so called parafermionic hyperbolic gamma functions \cite{Imamura:2012rq,Nieri:2015yia,Sarkissian:2018ppc}, defined for any $r\in \mathbb{N}$:
\bea \label{rrf} &&
\Lambda(y,m)=\prod_{k=0}^{m-1}S_b
\left({y\over r}+b\left(1-{m\over r}\right)+Q{k\over r}\right)\nonumber
\\ && \makebox[2em]{} \times
\prod_{k=0}^{r-m-1}S_b
\left({y\over r}+{m\over br}+Q{k\over r}\right)\, ,
\eea
where
$0\leq m\leq r$. Obviously, for $r=2$ one obtains the functions (\ref{nsf}) and (\ref{rf}):
\be\label{l1s}
\Lambda(y,0)= S_{1}(y),
\ee
\be\label{l2s}
\Lambda(y,1)= S_{0}(y)\, .
\ee
In paper \cite{Spiridonov:2016uae}, rarefied elliptic gamma functions are constructed and corresponding generalizations of the
elliptic beta integral and elliptic V-function are studied.
It was checked also in \cite{Sarkissian:2018ppc} that the functions (\ref{rrf}) are asymptotic forms of the
rarefied elliptic gamma functions. Therefore we can get the necessary identities using the same
chains of the reductions as in the case $r=1$. Some of the necessary integral identities were derived in this way in papers
\cite{Apresyan:2022erh,Sarkissian:2018ppc}. Others are obtained in appendices B-E .
Our main results are given by formulae \eqref{salbe} and \eqref{srul}. In fact these integrals
describe also bulk-boundary coupling in super Liouville theory as we explained above.

In super Liouville theory  we find as well that integrals describing matrix of modular transformation reside on the junction
of different theories.

First one can see that integral \eqref{salbe} is zero-holonomy value of
the partition function of the above mentioned mass-deformed $T[SU(2)]$ theory
on squashed lens space $S^3_b/\mathbb{Z}_2$ found in \cite{Nieri:2015yia}.
This is in agreement with AGT correspondence between $N=2$ $SU(2)$ supersymmetric gauge theory
on $S^4/\mathbb{Z}_2$ and super Liouville theory \cite{Belavin:2011pp,Belavin:2012aa,Belavin:2011tb,Bonelli:2011jx,Bonelli:2011kv}.
Thus we can consider our results as an evidence for generalization of the AGT correspondence between gauge theory
on  $S^4/\mathbb{Z}_2$ and super Liouville theory in the presence of the domain walls.
It is important further to study correspondence between holonomies entering in partition function of $T[SU(2)]$ theory
on squashed lens space $S^3_b/\mathbb{Z}_2$ and types of primaries entering in matrix of modular transformation.

The paper is organized as follows. In section 2, we review some basic facts on $N=1$ Super Liouville field theory.
In section 3, we calculate $S_{\beta_1\beta_2}(\beta_3)$ in the case when all $\beta_i$, $i=1,2,3$ are NS primaries.
The corresponding result is given by formulae \eqref{salbe} and \eqref{salbe2}.
In section 4, we calculate $S_{\beta_1\beta_2}(\beta_3)$ in the case when $\beta_i$, $i=1,2$ are NS primaries and $\beta_3$ is
\~{NS} primary. It is given by \eqref{srul} and \eqref{srulik}.
In section 5, we comment our results and outline possible applications.
In appendices A-E the necessary special functions,
their properties and integral identities are reviewed.


\section{Basics on $N=1$ super Liouville field theory}
$N=1$ super Liouville field theory is defined on a two-dimensional surface with metric $g_{ab}$ by the Lagrangian
density:
\be
{\cal L}={1\over 2\pi}g_{ab}\partial_a\varphi\partial_b \varphi+
{1\over 2\pi}(\psi\bar{\partial}\psi+\bar{\psi}\partial\bar{\psi})+2i\mu b^2\bar{\psi}\psi e^{b\varphi}+
2\pi \mu^2 b^2 e^{2b\varphi}\, .
\ee
The energy-momentum tensor and the superconformal current are
\bea
&&T=-{1\over 2} (\partial\varphi\partial\varphi-Q\partial^2\varphi+\psi\partial\psi)\, ,\\
&&G=i(\psi\partial\varphi-Q\partial\psi)\, ,
\eea
where $Q=b+{1\over b}$.
 The superconformal algebra is
 \bea
&& [L_m,L_n]=(m-n)L_{m+n}+{c_{SL}\over 12}m(m^2-1)\delta_{m+n}\, ,\\
 &&[L_m, G_k]={m-2k\over 2}G_{m+k}\, ,\\
 &&\{G_k,G_l\}=2L_{l+k}+{c\over 3}\left(k^2-{1\over 4}\right)\delta_{k+l}\, ,
 \eea
with the central charge
\be
c_{SL}={3\over 2}+3Q^2\, .
\ee
Here $k$ and $l$ take integer values for the Ramond algebra and half-integer values for the Neveu-Schwarz algebra.
Since in the Neveu-Schwarz sector of $N=1$ SLFT one has besides the Virasoro generators  $L_m$, also supercurrent generators $G_k$ with
half-integer $k$, descendant fields are broken into two sets of integer and half-integer levels.
Thus we will work with primary fields associated with the vertex operators
$N_{\alpha}=e^{\alpha \varphi}$, which we call NS field, and its supercurrent descendant $\tilde{N}_{\alpha}=G_{-1/2}N_{\alpha}$,
which we call  \~{NS} field \cite{Belavin:2007gz}.
The $N_{\alpha}$ primary field has conformal dimension
\be\label{hw1}
\Delta_{\alpha}={1\over 2}\alpha(Q-\alpha)\, .
\ee
The physical states have $\alpha={Q\over 2}+iP$.
The $\tilde{N}_{\alpha}$ has the conformal dimension
\be\label{hw2}
\Delta_{\tilde{\alpha}}={1\over 2}\alpha(Q-\alpha)+{1\over 2}\, .
\ee
The degenerate states are given by the momenta:
\be\label{degnm}
\alpha_{m,n}={1\over 2b}(1-m)+{b\over 2}(1-n)
\ee
with even $m-n$ in the NS sector and odd $m-n$ in the R sector.
For future use we need matrix of modular transformation of the NS characters.
The corresponding characters for generic $P$ which have no null-states are
\be\label{charl}
\chi_{P}^{NS}(\tau)=\sqrt{\theta_3(q)\over \eta(q)}{q^{P^2/2}\over \eta(\tau)}\, ,
\ee
where $q=\exp(2\pi i\tau)$ and
\be
\eta(\tau)=q^{1/24}\prod_{n=1}^{\infty}(1-q^n)\, .
\ee
Modular transformation of characters (\ref{charl}) is well-known:
\be\label{motrp1}
\chi_{P}^{NS}(\tau)=\int\chi_{P'}^{NS}(-1/\tau)e^{-2i\pi PP'}dP'\, .
\ee

For the degenerate representations, the characters are given by those of the corresponding Verma modules subtracted
by those of null submodules:
\be\label{deg1}
\chi_{m,n}^{NS}=\chi_{{1\over 2}(nb+mb^{-1})}^{NS}-\chi_{{1\over 2}(nb-mb^{-1})}^{NS}\, .
\ee
Modular transformations of (\ref{deg1})  is
\be\label{motrdi1}
\chi_{m,n}^{NS}(\tau)=\int\chi_{P}^{NS}(-1/\tau)2\sinh(\pi mP/b)\sinh(\pi nbP)dP\, .
\ee
Given that the identity field is specified by $(m,n)=(1,1)$  one finds the vacuum component of the matrix of modular transformation:
\be\label{soa}
S_{0\alpha}=\sin\pi b^{-1}(\alpha-Q/2)\sin\pi b(\alpha-Q/2)\, .
\ee
For brevity the $N_{\alpha}$ and $\tilde{N}_{\alpha}$ primary fields will be denoted by $\alpha$ and $\tilde{\alpha}$ correspondingly.
Therefore, if for example we take in the relation (\ref{behr}) $\alpha$, $\beta_2$ and $\beta_5$ of NS type, it takes the form

\bea\label{sabg}
&&S_{0\beta_2}\Bigg[\int {d\beta_4\over i}F_{0,\beta_4}\left[\begin{array}{cc}
\beta_2&\alpha\\
\beta_2&\alpha \end{array}\right]e^{2\pi i \Delta_{\beta_4}}
F_{\beta_4,\beta_5}\left[\begin{array}{cc}
\alpha&\alpha\\
\beta_2&\beta_2 \end{array}\right]\\ \nonumber
&&+\int {d\beta_4\over i}F_{0,\tilde{\beta_4}}\left[\begin{array}{cc}
\beta_2&\alpha\\
\beta_2&\alpha \end{array}\right]e^{2\pi i \Delta_{\tilde{\beta_4}}}
F_{\tilde{\beta_4},\beta_5}\left[\begin{array}{cc}
\alpha&\alpha\\
\beta_2&\beta_2 \end{array}\right]\Bigg]=\\ \nonumber
&&e^{\pi i (2\Delta_{\alpha}+2\Delta_{\beta_2})}
F_{0,\beta_5}\left[\begin{array}{cc}
\alpha&\alpha\\
\alpha&\alpha \end{array}\right]S_{\alpha\beta_2}(\beta_5)\, .
\eea
As we mentioned before $S_{\alpha\beta_2}(\beta_5)$ in the limit $\beta_5\to 0$ should coincide with the matrix of modular transformations
of the characters, and $S_{0\beta_2}$ is given by \eqref{soa}.

The elements of the fusion matrix in the Neveu-Schwarz sector of $N=1$ SLFT are given by the formulae \cite{Hadasz:2007wi,Hadasz:2013bwa}:
\be\label{fusik1}
 F_{\alpha_s,\alpha_t}\left[\begin{array}{cc}
\alpha_3& \alpha_2\\
\alpha_4& \alpha_1 \end{array}\right]={{\cal N}_{NS}(\alpha_s,\alpha_2,\alpha_1){\cal N}_{NS}(\alpha_4,\alpha_3,\alpha_s)
\over {\cal N}_{NS}(\alpha_t,\alpha_3,\alpha_2){\cal N}_{NS}(\alpha_4,\alpha_t,\alpha_1)}\big\{\,{}^{\alpha_1}_{\alpha_2}\,{}^{\alpha_3}_{{\alpha}_4}\,|\,{}^{\alpha_s}_{\alpha_t}\big\}_{11}^{11}\, ,
\ee
\be\label{fusik2}
 F_{\alpha_s,\tilde{\alpha_t}}\left[\begin{array}{cc}
\alpha_3& \alpha_2\\
\alpha_4& \alpha_1 \end{array}\right]={{\cal N}_{NS}(\alpha_s,\alpha_2,\alpha_1){\cal N}_{NS}(\alpha_4,\alpha_3,\alpha_s)
\over {\cal N}_{R}(\alpha_t,\alpha_3,\alpha_2){\cal N}_{R}(\alpha_4,\alpha_t,\alpha_1)}\big\{\,{}^{\alpha_1}_{\alpha_2}\,{}^{\alpha_3}_{{\alpha}_4}\,|\,{}^{\alpha_s}_{\alpha_t}\big\}_{00}^{11}\, ,
\ee

\be\label{fusik3}
 F_{\tilde{\alpha_s},\alpha_t}\left[\begin{array}{cc}
\alpha_3& \alpha_2\\
\alpha_4& \alpha_1 \end{array}\right]={{\cal N}_{R}(\alpha_s,\alpha_2,\alpha_1){\cal N}_{R}(\alpha_4,\alpha_3,\alpha_s)
\over {\cal N}_{NS}(\alpha_t,\alpha_3,\alpha_2){\cal N}_{NS}(\alpha_4,\alpha_t,\alpha_1)}\big\{\,{}^{\alpha_1}_{\alpha_2}\,{}^{\alpha_3}_{{\alpha}_4}\,|\,{}^{\alpha_s}_{\alpha_t}\big\}_{11}^{00}\, ,
\ee
\be\label{fusik4}
 F_{\tilde{\alpha_s},\tilde{\alpha_t}}\left[\begin{array}{cc}
\alpha_3& \alpha_2\\
\alpha_4& \alpha_1 \end{array}\right]={{\cal N}_{R}(\alpha_s,\alpha_2,\alpha_1){\cal N}_{R}(\alpha_4,\alpha_3,\alpha_s)
\over {\cal N}_{R}(\alpha_t,\alpha_3,\alpha_2){\cal N}_{R}(\alpha_4,\alpha_t,\alpha_1)}\big\{\,{}^{\alpha_1}_{\alpha_2}\,{}^{\alpha_3}_{{\alpha}_4}\,|\,{}^{\alpha_s}_{\alpha_t}\big\}_{00}^{00}\, ,
\ee
where normalization coefficients are described by the relations:
\bea\label{dozz1}
&&{\cal N}_{NS}(\alpha_3,\alpha_2,\alpha_1)=\\ \nonumber
&&{\Gamma_{NS}(2\alpha_1)\Gamma_{NS}(2\alpha_2)\Gamma_{NS}(2Q-2\alpha_3)\over
\Gamma_{NS}(2Q-\alpha_1-\alpha_2-\alpha_3)\Gamma_{NS}(Q-\alpha_1-\alpha_2+\alpha_3)
\Gamma_{NS}(\alpha_2+\alpha_3-\alpha_1)\Gamma_{NS}(\alpha_3+\alpha_1-\alpha_2)}\, ,
\eea
\bea\label{dozz2}
&&{\cal N}_{R }(\alpha_3,\alpha_2,\alpha_1)=\\ \nonumber
&&{\Gamma_{NS}(2\alpha_1)\Gamma_{NS}(2\alpha_2)\Gamma_{NS}(2Q-2\alpha_3)\over
\Gamma_{R}(2Q-\alpha_1-\alpha_2-\alpha_3)\Gamma_{R}(Q-\alpha_1-\alpha_2+\alpha_3)
\Gamma_{R}(\alpha_2+\alpha_3-\alpha_1)\Gamma_{R}(\alpha_3+\alpha_1-\alpha_2)}\, ,
\eea
where the functions $\Gamma_{NS}(x)$, $\Gamma_{R}(x)$ are described in appendix A, and $6j$-symbols of the quantum algebra ${\rm U}_q({\rm osp}(1|2))$ are \cite{Hadasz:2013bwa}:
\be\label{ponsotb}
\big\{\,{}^{\alpha_1}_{\alpha_3}\,{}^{\alpha_2}_{{\overline{\alpha}_4}}\,|\,{}^{\alpha_s}_{\alpha_t}\big\}_{\nu_1\nu_2}^{\nu_3\nu_4}=|S_1(2\alpha_t)|^2
{S_{\nu_4}(\alpha_s+\alpha_2-\alpha_1)S_{\nu_1}(\alpha_1+\alpha_t-\alpha_4)\over
S_{\nu_2}(\alpha_t+\alpha_2-\alpha_3)S_{\nu_3}(\alpha_3+\alpha_s-\alpha_4)}
I_{\alpha_s,\alpha_t}\left[\begin{array}{cc}
\alpha_3& \alpha_2\\
\alpha_4& \alpha_1 \end{array}\right]_{\nu_1\nu_2}^{\nu_3\nu_4},
\ee
where ${\overline{\alpha}_4}=Q-\alpha_4$, $\nu_i=0,1$, $i=1,2,3,4$ , $\sum_{i=1}^4\nu_i=0\, \, {\rm mod}\: 2$, and
\bea\label{rwhps}
&& I_{\alpha_s,\alpha_t}\left[\begin{array}{cc}
\alpha_3& \alpha_2\\
\alpha_4& \alpha_1 \end{array}\right]_{\nu_1\nu_2}^{\nu_3\nu_4}=(-1)^{\nu_3\nu_2+\nu_4}\int_{-\textup{i}\infty}^{\textup{i}\infty}\sum_{\nu=0}^{1}
(-1)^{\nu(\nu_2+\nu_4)}
S_{1+\nu+\nu_3}(y+\gamma_1^{\circ})\qquad\qquad
\\ \nonumber  && \makebox[2em]{} \times S_{1+\nu+\nu_4}(y+\gamma_2^{\circ})
S_{1+\nu+\nu_3}(y+\gamma_3^{\circ})S_{1+\nu+\nu_4}(y+\gamma_4^{\circ})
S_{\nu}(-y+\beta_1^{\circ})
\\ \nonumber && \makebox[2em]{} \times
S_{\nu+\nu_2+\nu_3}(-y+\beta_2^{\circ})S_{\nu+\nu_2+\nu_3}(-y+\beta_3^{\circ})
S_{\nu}(-y+\beta_4^{\circ})
{dy\over 2\textup{i}}\\ \nonumber
&&=\int_{-\textup{i}\infty}^{\textup{i}\infty}\sum_{\nu=0}^{1}
\Lambda(y+\gamma_1^{\circ},\nu_3+\nu)\Lambda(y+\gamma_2^{\circ},\nu+\nu_4)
\Lambda(y+\gamma_3^{\circ},\nu+\nu_3)
\\ \nonumber &&  \makebox[2em]{} \times
\Lambda(y+\gamma_4^{\circ},\nu-\nu_4)
\Lambda(-y+\beta_1^{\circ},-\nu-1)\Lambda(-y+\beta_2^{\circ}, -\nu+1+\nu_2-\nu_3)
\\  \nonumber &&  \makebox[2em]{} \times
\Lambda(-y+\beta_3^{\circ},-\nu+1-\nu_2-\nu_3)
\Lambda(-y+\beta_4^{\circ},-\nu-1){dy\over 2\textup{i}}\, ,
\eea
where
\be\label{ptparam}
\begin{aligned}
&\gamma_1^{\circ}=-Q/2+\alpha_3-\alpha_4\, ,\\
&\gamma_2^{\circ}=-Q/2+\alpha_1-\alpha_2\, ,\\
&\gamma_3^{\circ}=Q/2-\alpha_3-\alpha_4\, ,\\
&\gamma_4^{\circ}=Q/2-\alpha_1-\alpha_2\, ,
\end{aligned}\qquad
\begin{aligned}
&\beta_1^{\circ}=Q/2+\alpha_s\, ,\\
&\beta_2^{\circ}=Q/2-\alpha_t+\alpha_4+\alpha_2\, ,\\
&\beta_3^{\circ}=-Q/2+\alpha_t+\alpha_4+\alpha_2\, ,\\
&\beta_4^{\circ}=3Q/2-\alpha_s\, .
\end{aligned}
\ee
The second equality in \eqref{rwhps} was established in \cite{Apresyan:2022erh}.
Let us explain briefly how it arises. First recall
 the following periodicity and reflection properties of
$\Lambda(y,\nu)$\cite{Sarkissian:2018ppc,Sarkissian:2019bmt}:
\be\label{lambdasign2}
\Lambda(y,m+2k)=(-1)^{mk+r{k(k-1)\over 2}}\Lambda(y,m)\, ,
\ee
\be\label{lambdasign3}
\Lambda(Q-x,-\nu)\Lambda(x,\nu)=(-1)^{\nu}\, .
\ee
For $r=2$ the relation \eqref{lambdasign2} implies
\be\label{lambdasign23}
\Lambda(y,m+2k)=(-1)^{mk}\Lambda(y,m)\, .
\ee
The functions $S_{\nu}(x)$ are supposed to be periodic in the discrete variable: $S_{\nu}(x)=S_{\nu+2}(x)$.
This together with \eqref{lambdasign23} implies
\be\label{l1s2}
\Lambda(y,2m)=
  S_{1}(y),
\quad
\Lambda(y,2m+1)=(-1)^{m}
 S_{0}(y)\, .
\ee
Repeatedly using \eqref{l1s2} one  can arrive to the second  equality in \eqref{rwhps}.
One can check that the expressions \eqref{fusik1}-\eqref{fusik4} are invariant under the reflection $\alpha\to Q-\alpha$ of any
of six variables entering in the fusion matrix.
Inserting \eqref{fusik1}-\eqref{fusik4} in \eqref{sabg} we derive:
\be
S^{SL}_{\alpha\beta_2}(\beta_5)={{\cal N}_{NS}(\alpha,\beta_5,\alpha)\over {\cal N}_{NS}(\beta_2,\beta_5,\beta_2)}S_{\alpha\beta_2}(\beta_5)
\ee
where $S_{\alpha\beta_2}(\beta_5)$ is computed by the same formula \eqref{sabg} where inserted only corresponding elements
of $6j$-symbols of the quantum algebra ${\rm U}_q({\rm osp}(1|2))$ appearing in the l.h.s of \eqref{fusik1}-\eqref{fusik4}:
\be\label{sabgg3}
S_{0\beta_2}I_{\alpha\beta_2}(\beta_5)=e^{\pi i (2\Delta_{\alpha}+2\Delta_{\beta_2})}
\big\{\,{}^{\alpha}_{\alpha}\,{}^{\alpha}_{\alpha}\,|\,{}^{0}_{\beta_5}\big\}_{11}^{11}
S_{\alpha\beta_2}(\beta_5)\, ,
\ee
where
\bea\label{sabgg}
&&I_{\alpha\beta_2}(\beta_5)=\Bigg[\int {d\beta_4\over i}\big\{\,{}^{\alpha}_{\alpha}\,{}^{\beta_2}_{\beta_2}\,|\,{}^{0}_{\beta_4}\big\}_{11}^{11}e^{2\pi i \Delta_{\beta_4}}\big\{\,{}^{\beta_2}_{\alpha}\,{}^{\alpha}_{\beta_2}\,|\,{}^{\beta_4}_{\beta_5}\big\}_{11}^{11}
\\ \nonumber
&&+\int {d\beta_4\over i}\big\{\,{}^{\alpha}_{\alpha}\,{}^{\beta_2}_{\beta_2}\,|\,{}^{0}_{\beta_4}\big\}_{00}^{11}e^{2\pi i \Delta_{\tilde{\beta_4}}}\big\{\,{}^{\beta_2}_{\alpha}\,{}^{\alpha}_{\beta_2}\,|\,{}^{\beta_4}_{\beta_5}\big\}_{11}^{00}
\Bigg]\, .
\eea

\section{Elements of S-matrix for three NS primaries}

Now we turn to the calculation of the matrix $S_{\alpha\beta_2}(\beta_5)$  in the case when all the entering primaries of the NS type.
For this purpose we use the formula (\ref{sabgg3}). Start to calculate the left hand side integral \eqref{sabgg}. The first term in   (\ref{sabgg}) is the following element of the $6j$-symbol: $\big\{\,{}^{\alpha}_{\alpha}\,{}^{\beta_2}_{\beta_2}\,|\,{}^{0}_{\beta_4}\big\}_{11}^{11}$.

Setting in (\ref{ponsotb})-\eqref{rwhps}
$\nu_1=\nu_2=\nu_3=\nu_4=1$, and $\alpha_1=\alpha_2=\alpha$,     $\alpha_3=\beta_2$,   $\alpha_4=Q-\beta_2$, $\alpha_t=\beta_4$, $\alpha_s=\delta\to 0$ we obtain
\bea\label{ponsott}
&&\big\{\,{}^{\alpha}_{\alpha}\,{}^{\beta_2}_{\beta_2}\,|\,{}^{\delta}_{\beta_4}\big\}_{11}^{11}=|S_1(2\beta_4)|^2
{S_{1}(\delta)S_{1}(\alpha+\beta_4+\beta_2-Q)\over
S_{1}(\beta_4+\alpha-\beta_2)S_{1}(2\beta_2-Q)}\\ \nonumber
&&\times\int_{-\textup{i}\infty}^{\textup{i}\infty}\sum_{\nu=0}^{1}(-1)^{\nu+1}
\Lambda(y-3Q/2+2\beta_2,1+\nu)
\Lambda(y-Q/2,\nu+1)\Lambda(y+Q/2-2\alpha,\nu-1)\qquad
\\ \nonumber &&  \times
\Lambda(-y+Q/2+\delta,-\nu-1)\Lambda(-y+3Q/2-\beta_4-\beta_2+\alpha, -\nu+1)\\ \nonumber
  &&  \times
\Lambda(-y+Q/2+\beta_4-\beta_2+\alpha,-\nu-1){dy\over 2i}\\
\nonumber
&&=|S_1(2\beta_4)|^2
S_{1}(\delta)S_{1}(\alpha+\beta_4+\beta_2-Q)
\Lambda(\delta,0)
\Lambda(Q-\beta_2-\beta_4+\alpha,2)\Lambda(-2\alpha,-2)\, .
\eea
To compute integral here we used formula (\ref{pf2}) in appendix B for $r=2$.

Now we should compute the second term in (\ref{sabgg}): $\big\{\,{}^{\beta_2}_{\alpha}\,{}^{\alpha}_{\beta_2}\,|\,{}^{\beta_4}_{\beta_5}\big\}_{11}^{11}$.
Setting in (\ref{ponsotb})
$\nu_1=\nu_2=\nu_3=\nu_4=1$, and $\alpha_1=\beta_2$,  $\alpha_2=\alpha$,     $\alpha_3=\alpha$,   $\alpha_4=Q-\beta_2$, $\alpha_t=\beta_5$, $\alpha_s=\beta_4$ we have
\bea\label{qonoott}
&&\big\{\,{}^{\beta_2}_{\alpha}\,{}^{\alpha}_{\beta_2}\,|\,{}^{\beta_4}_{\beta_5}\big\}_{11}^{11}=|S_1(2\beta_5)|^2
{S_{1}(\alpha+\beta_4-\beta_2)S_{1}(2\beta_2+\beta_5-Q)\over
S_{1}(\beta_4+\alpha+\beta_2-Q)S_{1}(\beta_5)}\\ \nonumber
&&\times\int_{-\textup{i}\infty}^{\textup{i}\infty}\sum_{\nu=0}^{1}
\Lambda(-y+Q/2+\beta_4,-\nu-1)\Lambda(-y+3Q/2-\beta_4,-\nu-1)F(y,\nu+1)(-1)^{\nu-1}{dy\over 2i}\, ,\\
\nonumber
\eea
where
\bea\label{auf}
&&F(y,\nu)=\Lambda(y-3Q/2+\beta_2+\alpha, \nu)\Lambda(y-Q/2+\beta_2-\alpha,\nu)\\ \nonumber &&  \times
\Lambda(y-Q/2-\alpha+\beta_2,\nu)\Lambda(y+Q/2-\alpha-\beta_2,\nu)\\ \nonumber &&  \times
\Lambda(-y+3Q/2-\beta_5-\beta_2+\alpha, -\nu+2)\Lambda(-y+Q/2+\beta_5-\beta_2+\alpha,-\nu)\, .
\eea
Writing in this way we separated in integral appearing in (\ref{qonoott}) $\beta_4$-dependent terms and collected in
$F(y,\nu)$ all $\beta_4$-independent terms.
The next element of the fusion matrix to be computed is $\big\{\,{}^{\alpha}_{\alpha}\,{}^{\beta_2}_{\beta_2}\,|\,{}^{\delta}_{\beta_4}\big\}_{00}^{11}$.
Putting in (\ref{ponsotb})
$\nu_1=\nu_2=0$, $\nu_3=\nu_4=1$, and $\alpha_1=\alpha_2=\alpha$,     $\alpha_3=\beta_2$,   $\alpha_4=Q-\beta_2$, $\alpha_t=\beta_4$, $\alpha_s=\delta\to 0$ one has
\bea\label{donsott}
&&\big\{\,{}^{\alpha}_{\alpha}\,{}^{\beta_2}_{\beta_2}\,|\,{}^{\delta}_{\beta_4}\big\}_{00}^{11}=|S_1(2\beta_4)|^2
{S_{1}(\delta)S_{0}(\alpha+\beta_4+\beta_2-Q)\over
S_{0}(\beta_4+\alpha-\beta_2)S_{1}(2\beta_2-Q)}\\ \nonumber
&&\times\int_{-\textup{i}\infty}^{\textup{i}\infty}\sum_{\nu=0}^{1}(-1)^{\nu+1}
\Lambda(y-3Q/2+2\beta_2,1+\nu)
\Lambda(y-Q/2,\nu+1)\Lambda(y+Q/2-2\alpha,\nu-1)\qquad
\\ \nonumber &&  \times
\Lambda(-y+Q/2+\delta,-\nu-1)\Lambda(-y+3Q/2-\beta_4-\beta_2+\alpha, -\nu)
\Lambda(-y+Q/2+\beta_4-\beta_2+\alpha,-\nu){dy\over 2i}\\
\nonumber
&&=|S_1(2\beta_4)|^2
S_{1}(\delta)S_{0}(\alpha+\beta_4+\beta_2-Q)
\Lambda(\delta,0)
\Lambda(Q-\beta_2-\beta_4+\alpha,1)\Lambda(-2\alpha,-2)\, .
\eea

The last element of the fusion matrix which we need to compute is $\big\{\,{}^{\beta_2}_{\alpha}\,{}^{\alpha}_{\beta_2}\,|\,{}^{\beta_4}_{\beta_5}\big\}_{11}^{00}$.
Now setting  in (\ref{ponsotb})
$\nu_1=\nu_2=1$, $\nu_3=\nu_4=0$, and $\alpha_1=\beta_2$,  $\alpha_2=\alpha$,     $\alpha_3=\alpha$,   $\alpha_4=Q-\beta_2$, $\alpha_t=\beta_5$, $\alpha_s=\beta_4$ one gets:

\bea\label{ponsottqq}
&&\big\{\,{}^{\beta_2}_{\alpha}\,{}^{\alpha}_{\beta_2}\,|\,{}^{\beta_4}_{\beta_5}\big\}_{11}^{00}=|S_1(2\beta_5)|^2
{S_{0}(\alpha+\beta_4-\beta_2)S_{1}(2\beta_2+\beta_5-Q)\over
S_{0}(\beta_4+\alpha+\beta_2-Q)S_{1}(\beta_5)}\\ \nonumber
&&\times\int_{-\textup{i}\infty}^{\textup{i}\infty}\sum_{\nu=0}^{1}
\Lambda(-y+Q/2+\beta_4,-\nu-1)\Lambda(-y+3Q/2-\beta_4,-\nu-1)F(y,\nu){dy\over 2i}\, ,\\
\nonumber
\eea

were the function $F(y,\nu)$ was defined in (\ref{auf}).

Since $\beta_4$ has the form $\beta_4=Q/2+iP$ for some real $P$, one can write
\bea\nonumber
&&|S_1(2\beta_4)|^2=S_1(2\beta_4)S_1(2Q-2\beta_4)={1\over S_1(Q-2\beta_4)S_1(-Q+2\beta_4)}\\ \nonumber
&&={1\over S_1(2(Q/2-\beta_4))S_1(2(-Q/2+\beta_4))}={1\over S_1(2z)S_1(-2z)}\, ,\nonumber
\eea
where  $z=-Q/2+\beta_4$. Recalling also formulae for conformal dimensions (\ref{hw1}) and (\ref{hw2}) we obtain

\bea\label{cczz}
&&I_{\alpha\beta_2}(\beta_5)\\ \nonumber
&&=C\int_{-\textup{i}\infty}^{\textup{i}\infty}{dy\over i}F(y,0)\int_{0}^{\textup{i}\infty} {dz\over 2i}\sum_{m=0}^{1}{\Lambda(Q/2+\alpha-\beta_2\pm z,\pm m)\Lambda(-y+Q\pm z, \pm m)\over \Lambda(\pm 2z ,\pm 2m)}e^{-i\pi z^2}(-1)^m\\ \nonumber
&&-C\int_{-\textup{i}\infty}^{\textup{i}\infty}{dy\over i}F(y,1)\int_{0}^{\textup{i}\infty} {dz\over 2i}\sum_{m=0}^{1}{\Lambda(Q/2+\alpha-\beta_2\pm z,1\pm m)\Lambda(-y+Q\pm z, \pm m)\over \Lambda(\pm 2z ,\pm 2m)}e^{-i\pi z^2}(-1)^m\, ,
\eea
\be
z=\beta_4-{Q\over 2}\, ,\qquad C=2S_{1}(\delta)\Lambda(\delta,0)\Lambda(-2\alpha,-2){|S_1(2\beta_5)|^2S_{1}(2\beta_2+\beta_5-Q)\over
S_{1}(\beta_5)}e^{Q^2\pi i/4}\, .
\ee
Here we used the shorthand notation $\Lambda(y\pm x ,m\pm n)\equiv\Lambda(y+ x ,m+ n)\Lambda(y- x ,m- n)$. The integrals appearing in the third and the fourth lines of (\ref{cczz}) are particular cases of the  integral (\ref{pf1})  in appendix B for $r=2$.
Hence using  (\ref{pf1}) in appendix B for $r=2$ one has

\bea\label{ccinter}
&&I_{\alpha\beta_2}(\beta_5)
=C\int_{-\textup{i}\infty}^{\textup{i}\infty}{dy\over i}F(y,0)\Lambda(3Q/2+\alpha-\beta_2-y,0)
e^{-{i\pi\over 4}(-Q/2+\alpha-\beta_2+y)^2}e^{{i\pi\over 4}(3Q/2+\alpha-\beta_2-y)Q}\qquad\\ \nonumber
&&-C\int_{-\textup{i}\infty}^{\textup{i}\infty}{dy\over i}F(y,1)\Lambda(3Q/2+\alpha-\beta_2-y,1)
e^{-{i\pi\over 4}(-Q/2+\alpha-\beta_2+y)^2}e^{{i\pi\over 4}(3Q/2+\alpha-\beta_2-y)Q}e^{-{i\pi\over 4}}e^{{i\pi\over 2}}\\ \nonumber
&&=C'\int_{-\textup{i}\infty}^{\textup{i}\infty}{dy\over i}\sum_{\nu=0}^{1}
\Lambda(y-3Q/2+\beta_2+\alpha, \nu)\Lambda(y-Q/2-\alpha+\beta_2,\nu)\Lambda(y+Q/2-\alpha-\beta_2,\nu)\qquad\\ \nonumber &&  \times
\Lambda(-y+3Q/2-\beta_5-\beta_2+\alpha, -\nu+2)\Lambda(-y+Q/2+\beta_5-\beta_2+\alpha,-\nu)\\ \nonumber
&&\times e^{-{i\pi\over 4}y^2}e^{-{i\pi\over 2}y(\alpha-\beta_2)}e^{-{i\pi\nu\over 4}}e^{{3i\pi\nu\over 2}}\, ,
\eea
where

\be
C'=Ce^{-{i\pi\over 4}(-Q/2+\alpha-\beta_2)^2}e^{{i\pi\over 4}(3Q/2+\alpha-\beta_2)Q}\, .
\ee
Changing sign of $y$ and using that for $\nu=0,1$ hold the relations: $\Lambda(y, -\nu)=(-1)^{\nu}\Lambda(y, \nu)$ and
$\Lambda(y, 2-\nu)=\Lambda(y, \nu)$ we can write the last integral in (\ref{ccinter}) in the form
\bea\nonumber
&&I_{\alpha\beta_2}(\beta_5)=C'\int_{-\textup{i}\infty}^{\textup{i}\infty}{dy\over i}\sum_{\nu=0}^{1}
\Lambda(-y-3Q/2+\beta_2+\alpha, -\nu)\Lambda(-y-Q/2-\alpha+\beta_2,-\nu)\qquad\\ \nonumber
 &&  \times
\Lambda(-y+Q/2-\alpha-\beta_2,-\nu)\Lambda(y+3Q/2-\beta_5-\beta_2+\alpha, \nu)\Lambda(y+Q/2+\beta_5-\beta_2+\alpha,\nu)\\ \nonumber
&&\times e^{-{i\pi\over 4}y^2}e^{{i\pi\over 2}y(\alpha-\beta_2)}e^{-{i\pi\nu\over 4}}e^{{3i\pi\nu\over 2}}\, .
\eea

It is easy to see that this integral has the same structure as the integral on the left hand side of relation ({\ref{pf5}) in appendix D for $r=2$, therefore we can write
for him taking as $\gamma_1=Q/2+\beta_5-\beta_2+\alpha$ and $\gamma_2=3Q/2-\beta_5-\beta_2+\alpha$

\bea\label{bbvv}
&&I_{\alpha\beta_2}(\beta_5)=C'e^{{i\pi \over 4}((\alpha-\beta_2)^2+Q^2/4-Q(Q/2+\beta_5+2\alpha-2\beta_2))}\\ \nonumber
&&\times\Lambda(2\alpha-\beta_5,0)\Lambda(Q-\beta_5,0)\Lambda(2Q-2\beta_2-\beta_5,0)\\ \nonumber
&&\times \int_{-i\infty}^{i\infty}\sum_{m=0}^{1}{\Lambda(\pm y-Q+\beta_2+\alpha+\beta_5/2, \pm m)
\Lambda(\pm y-\alpha+\beta_2+\beta_5/2,\pm m)
\over \Lambda(2y,2m)\Lambda(-2y,-
(2m))}\\ \nonumber
&&\times\Lambda(\pm y+Q-\alpha-\beta_2+\beta_5/2, \pm m)
\Lambda(\pm y+\alpha-\beta_2+\beta_5/2,\pm m)e^{-i\pi y^2}
e^{-i\pi m^2}
{dy\over 2i}\, .
\eea

Note that integral in the r.h.s. of (\ref{bbvv}) coincides with the integral in the r.h.s. of (\ref{pf4}) in appendix C for $r=2$. Thus we can write for the latter

\bea\nonumber
&&\int_{-i\infty}^{i\infty}\sum_{m=0}^{1}\Lambda(\pm y-Q/2+\beta_2+\beta_5/2, \pm m)\Lambda(\pm y+Q/2-\beta_2+\beta_5/2,\pm m) e^{2i\pi y(-Q/2+\alpha)}  {dy\over i}
=\\ \nonumber
&&\Lambda(2\alpha-\beta_5,0)
\Lambda(-2\alpha-\beta_5+2Q,0)
e^{-i\pi\left({1\over 4}\beta_5^2-(-Q/2+\alpha)^2-(Q/2-\beta_2)^2\right)}\\ \nonumber
&&\times \int_{-i\infty}^{i\infty}\sum_{m=0}^{1}{\Lambda(\pm y-Q+\beta_2+\alpha+\beta_5/2, \pm m)
\Lambda(\pm y-\alpha+\beta_2+\beta_5/2,\pm m)
\over \Lambda(2y,2m)\Lambda(-2y,-
(2m))}\\ \nonumber
&&\times\Lambda(\pm y+Q-\alpha-\beta_2+\beta_5/2, \pm m)
\Lambda(\pm y+\alpha-\beta_2+\beta_5/2,\pm m)e^{-i\pi y^2}
e^{-i\pi m^2}
{dy\over 2i}\, .
\eea
Collecting all we obtain:
\bea\label{salbe}
&&S_{\alpha\beta_2}(\beta_5)=S_{0\beta_2}{e^{-{i\pi\over 2}\Delta_{\beta_5}}\over S_1(\beta_5)}\quad\\ \nonumber
&&\times\int_{-i\infty}^{i\infty}\sum_{m=0}^{1}\Lambda(\pm y-Q/2+\beta_2+\beta_5/2,\pm  m)
\Lambda(\pm y+Q/2-\beta_2+\beta_5/2, \pm m)e^{2i\pi y(-Q/2+\alpha)}  {dy\over i}
\eea
or using \eqref{l1s2}:
\bea\label{salbe2}
&&S_{\alpha\beta_2}(\beta_5)=S_{0\beta_2}{e^{-{i\pi\over 2}\Delta_{\beta_5}}\over S_{\rm NS}(\beta_5)}\quad\\ \nonumber
&&\times\int_{-i\infty}^{i\infty}\Big[S_{\rm NS}(\pm y-Q/2+\beta_2+\beta_5/2)S_{\rm NS}(\pm y+Q/2-\beta_2+\beta_5/2)\\ \nonumber
&&+S_{\rm R}(\pm y-Q/2+\beta_2+\beta_5/2)S_{\rm R}(\pm y+Q/2-\beta_2+\beta_5/2)\Big]
e^{2i\pi y(-Q/2+\alpha)}  {dy\over i}\, .
\eea
This expression has number of properties analogous to that of in usual Liouville field theory  \cite{Vartanov:2013ima}.
First we obtain easily
\be\label{sabo1}
S_{\alpha\beta_2}(\beta_5)=S_{Q-\alpha,\beta_2}(\beta_5)=S_{\alpha, Q-\beta_2}(\beta_5)\, .
\ee
Using formula \eqref{symref} in appendix E for $r=2$ one can establish
\bea\label{sabo2}
&&S_{\alpha\beta_2}(Q-\beta_5)=S_{\rm NS}(-Q+2\alpha+\beta_5)S_{\rm NS}(Q-2\alpha+\beta_5)\\ \nonumber
&&\times S_{\rm NS}(2Q-\beta_5-2\beta_2)S_{\rm NS}(-\beta_5+2\beta_2)
S_{\alpha\beta_2}(\beta_5)\, .
\eea
Combining \eqref{sabo1} and \eqref{sabo2} with definition of the normalization factors \eqref{dozz1} and using formulae \eqref{lsns} in appendix A we obtain that
\be
S^{SL}_{\alpha\beta_2}(\beta_5)={{\cal N}_{NS}(\alpha,\beta_5,\alpha)\over {\cal N}_{NS}(\beta_2,\beta_5,\beta_2)}S_{\alpha\beta_2}(\beta_5)
\ee
 is invariant under reflection of all three variables.

Now let us check that \eqref{salbe2} is self-consistent in the limit $\beta_5\to 0$. In this limit using formulae in appendix A
one has that ${1\over S_{\rm NS}(\beta_5)}\sim \pi \beta_5\to 0$, and thus we should pick up only diverging part of integral.
This diverging part originates due to pinching of integration contour in this limit between  poles of
$S_{\rm NS}(y-Q/2+\beta_2+\beta_5/2)$ and $S_{\rm NS}(-y+Q/2-\beta_2+\beta_5/2)$ and also between  poles of
$S_{\rm NS}(-y-Q/2+\beta_2+\beta_5/2)$ and $S_{\rm NS}(y+Q/2-\beta_2+\beta_5/2)$. Analyzing poles of  $S_R(x)$
reviewed in appendix A we see that the second term in  \eqref{salbe2}
has no diverging part. The diverging part caused by pinching is pole with coefficient given by the sum of residues:
\bea\nonumber
&&{2\pi i\over i\pi^2 \beta_5}S_{\rm NS}(-Q+2\beta_2)S_{\rm NS}(Q-2\beta_2)(e^{2i\pi (Q/2-\beta_2)(-Q/2+\alpha)}+e^{-2i\pi (Q/2-\beta_2)(-Q/2+\alpha)})\\ \nonumber
&&=-{1\over \pi \beta_5}{\cos\pi (Q/2-\beta_2)(-Q/2+\alpha)\over \sin{\pi\over b}(Q/2-\beta_2)\sin\pi b(Q/2-\beta_2)}\, ,
\eea
where the product $S_{\rm NS}(-Q+2\beta_2)S_{\rm NS}(Q-2\beta_2)$ was computed using formulae \eqref{kazo} and
 \eqref{kazol} in appendix A. Recalling \eqref{soa} and \eqref{motrp1} we obtain that indeed
\be
\stackreb{\lim}{\beta_5\to 0}S^{SL}_{\alpha\beta_2}(\beta_5)=S_{\alpha\beta_2}\, .
\ee

\section{Elements of S-matrix for two NS and one \~{NS} primaries}
Here we consider the case when $\alpha,\beta_2$ are NS primaries and $\beta_5$ is \~{NS} primary. In this case the relation (\ref{sabg}) takes the form
\be
S^{SL}_{\alpha\beta_2}(\tilde{\beta_5})={{\cal N}_{R}(\alpha,\beta_5,\alpha)\over {\cal N}_{R}(\beta_2,\beta_5,\beta_2)}S_{\alpha\beta_2}(\tilde{\beta_5})\, ,
\ee
where $S_{\alpha\beta_2}(\tilde{\beta_5})$ is given by the relation:
\be
S_{0\beta_2}I_{\alpha\beta_2}(\tilde{\beta_5})=e^{\pi i (2\Delta_{\alpha}+2\Delta_{\beta_2})}
\big\{\,{}^{\alpha}_{\alpha}\,{}^{\alpha}_{\alpha}\,|\,{}^{\delta}_{\beta_5}\big\}_{00}^{11}\;
S_{\alpha\beta_2}(\tilde{\beta_5})
\ee
and
\bea\label{sabgg2}
&&I_{\alpha\beta_2}(\tilde{\beta_5})=\Bigg[\int {d\beta_4\over i}\big\{\,{}^{\alpha}_{\alpha}\,{}^{\beta_2}_{\beta_2}\,|\,{}^{\delta}_{\beta_4}\big\}_{11}^{11}\;e^{2\pi i \Delta_{\beta_4}}\;\big\{\,{}^{\beta_2}_{\alpha}\,{}^{\alpha}_{\beta_2}\,|\,{}^{\beta_4}_{\beta_5}\big\}_{00}^{11}
\\ \nonumber
&&+\int {d\beta_4\over i}\big\{\,{}^{\alpha}_{\alpha}\,{}^{\beta_2}_{\beta_2}\,|\,{}^{\delta}_{\beta_4}\big\}_{00}^{11}\;e^{2\pi i \Delta_{\tilde{\beta_4}}}\;\big\{\,{}^{\beta_2}_{\alpha}\,{}^{\alpha}_{\beta_2}\,|\,{}^{\beta_4}_{\beta_5}\big\}_{00}^{00}
\Bigg]\, .
\eea

Here $\delta$ should be taken to $0$. The first terms in the first and second lines $\big\{\,{}^{\alpha}_{\alpha}\,{}^{\beta_2}_{\beta_2}\,|\,{}^{\delta}_{\beta_4}\big\}_{11}^{11}$
and $\big\{\,{}^{\alpha}_{\alpha}\,{}^{\beta_2}_{\beta_2}\,|\,{}^{\delta}_{\beta_4}\big\}_{00}^{11}$
are the same as before and computed in (\ref{ponsott}) and (\ref{donsott}) respectively.

To compute the second fusion matrix element in the first line we set  in (\ref{ponsotb})
$\nu_1=\nu_2=0$,  $\nu_3=\nu_4=1$, and $\alpha_1=\beta_2$,  $\alpha_2=\alpha$,     $\alpha_3=\alpha$,   $\alpha_4=Q-\beta_2$, $\alpha_t=\beta_5$, $\alpha_s=\beta_4$ and receive
\bea\label{lonsottq}
&&\big\{\,{}^{\beta_2}_{\alpha}\,{}^{\alpha}_{\beta_2}\,|\,{}^{\beta_4}_{\beta_5}\big\}_{00}^{11}=|S_1(2\beta_5)|^2
{S_{1}(\alpha+\beta_4-\beta_2)S_{0}(2\beta_2+\beta_5-Q)\over
S_{1}(\beta_4+\alpha+\beta_2-Q)S_{0}(\beta_5)}\\ \nonumber
&&\times\int_{-\textup{i}\infty}^{\textup{i}\infty}\sum_{\nu=0}^{1}
\Lambda(-y+Q/2+\beta_4,-\nu-1)\Lambda(-y+3Q/2-\beta_4,-\nu-1)F_2(y,\nu+1)(-1)^{\nu-1}{dy\over 2i}\, ,\\
\nonumber
\eea
where we defined
\bea\label{f2yn}
&&F_2(y,\nu)=\Lambda(y-3Q/2+\beta_2+\alpha,\nu)\Lambda(y-Q/2+\beta_2-\alpha,\nu)\\ \nonumber &&  \times
\Lambda(y-Q/2-\alpha+\beta_2,\nu)\Lambda(y+Q/2-\alpha-\beta_2,\nu)\\ \nonumber &&  \times
\Lambda(-y+3Q/2-\beta_5-\beta_2+\alpha, -\nu+1)
\Lambda(-y+Q/2+\beta_5-\beta_2+\alpha,-\nu+1)\, .
\eea

To derive  the second fusion matrix element in the second line
we set  in (\ref{ponsotb}):
$\nu_1=\nu_2=0$,  $\nu_3=\nu_4=0$, and $\alpha_1=\beta_2$,  $\alpha_2=\alpha$,     $\alpha_3=\alpha$,   $\alpha_4=Q-\beta_2$, $\alpha_t=\beta_5$, $\alpha_s=\beta_4$ and obtain

\bea\label{aonsottq}
&&\big\{\,{}^{\beta_2}_{\alpha}\,{}^{\alpha}_{\beta_2}\,|\,{}^{\beta_4}_{\beta_5}\big\}_{00}^{00}=|S_1(2\beta_5)|^2
{S_{0}(\alpha+\beta_4-\beta_2)S_{0}(2\beta_2+\beta_5-Q)\over
S_{0}(\beta_4+\alpha+\beta_2-Q)S_{0}(\beta_5)}\\ \nonumber
&&\times\int_{-\textup{i}\infty}^{\textup{i}\infty}\sum_{\nu=0}^{1}
\Lambda(-y+Q/2+\beta_4,-\nu-1)\Lambda(-y+3Q/2-\beta_4,-\nu-1)F_2(y,\nu){dy\over 2i}\, .\\
\nonumber
\eea
Collecting all, and repeating the same steps as in the first case and using the integral (\ref{pf1}) in appendix B, we obtain
\bea\label{intbtype}
&&I_{\alpha\beta_2}(\tilde{\beta}_5)=C''\int_{-\textup{i}\infty}^{\textup{i}\infty}{dy\over i}\sum_{\nu=0}^{1}
\Lambda(-y-3Q/2+\beta_2+\alpha, -\nu)\Lambda(-y-Q/2-\alpha+\beta_2,-\nu)\qquad\\ \nonumber
 &&  \times
\Lambda(-y+Q/2-\alpha-\beta_2,-\nu)\Lambda(y+3Q/2-\beta_5-\beta_2+\alpha, \nu+1)\\ \nonumber
&&\times \Lambda(y+Q/2+\beta_5-\beta_2+\alpha,\nu+1)e^{-{i\pi\over 4}y^2}e^{{i\pi\over 2}y(\alpha-\beta_2)}e^{-{i\pi\nu\over 4}}e^{{i\pi\nu\over 2}}\, ,
\eea
where
\bea
&&C''=S_{1}(\delta)\Lambda(\delta,0)\Lambda(-2\alpha,-2){|S_1(2\beta_5)|^2S_{0}(2\beta_2+\beta_5-Q)\over
S_{0}(\beta_5)}\\ \nonumber
&&\times e^{-{i\pi\over 4}(-Q/2+\alpha-\beta_2)^2}e^{{i\pi\over 4}(3Q/2+\alpha-\beta_2)Q}e^{Q^2\pi i/4}\, .
\eea

The integral (\ref{intbtype}) is of the form (\ref{pf5}) in appendix D, which yields
\bea\label{sif}
&&I_{\alpha\beta_2}(\tilde{\beta}_5)=C''e^{{i\pi \over 4}((\alpha-\beta_2)^2+Q^2/4-Q(Q/2+\beta_5+2\alpha-2\beta_2))}e^{i\pi\over 4}\\ \nonumber
&&\times\Lambda(2\alpha-\beta_5,1)\Lambda(Q-\beta_5,1)\Lambda(2Q-2\beta_2-\beta_5,1)\int_{-i\infty}^{i\infty}\sum_{m=0}^{1}\\ \nonumber
&&{\Lambda(\pm y-Q+\beta_2+\alpha+\beta_5/2, \pm m-\theta(\mp 1))
\Lambda(\pm y-\alpha+\beta_2+\beta_5/2,\pm m-\theta(\mp 1))
\over \Lambda(2y,2m+1)\Lambda(-2y,-
2m-1)}\\ \nonumber
&&\times\Lambda(\pm y+Q-\alpha-\beta_2+\beta_5/2, \pm m-\theta(\mp 1))
\Lambda(\pm y+\alpha-\beta_2+\beta_5/2,\pm m+1+\theta(\pm 1))\\ \nonumber
&&\times e^{-i\pi y^2}
{dy\over 2i}\, .
\eea
Here $\theta(x)$ is Heaviside step function: $\theta(-1)=0$ and $\theta(1)=1$. As before the functions with double sign in arguments
are used as shorthand for their product where the first term is taken with upper signs everywhere and the second term is taken with lower signs everywhere.
For example the first term in integrand reads
\bea
&&\Lambda(\pm y-Q+\beta_2+\alpha+\beta_5/2, \pm m-\theta(\mp 1))\equiv\\ \nonumber
&&\Lambda( y-Q+\beta_2+\alpha+\beta_5/2, m)
\Lambda(-y-Q+\beta_2+\alpha+\beta_5/2, -m-1)\, .
\eea
And finally we express the r.h.s. of (\ref{sif}) via (\ref{pf4}) in appendix C
\bea
&&\int_{-i\infty}^{i\infty}\sum_{m=0}^{1}\Lambda(y-Q/2+\beta_2+\beta_5/2, m-1)\Lambda(y+Q/2-\beta_2+\beta_5/2, m+1)\\ \nonumber
&&\Lambda(-y-Q/2+\beta_2+\beta_5/2,-m)\Lambda(-y+Q/2-\beta_2+\beta_5/2, -m)e^{2i\pi y(-Q/2+\alpha)} e^{i\pi m} {dy\over i}
=\\ \nonumber
&&\Lambda(2\alpha-\beta_5,1)
\Lambda(-2\alpha-\beta_5+2Q,-1)
e^{-i\pi\left({1\over 4}\beta_5^2-(\alpha-Q/2)^2-(Q/2-\beta_2)^2\right)}e^{i\pi\over 2}\int_{-i\infty}^{i\infty}\sum_{m=0}^{1}\\ \nonumber
&&{\Lambda(\pm y-Q+\beta_2+\alpha+\beta_5/2, \pm m-\theta(\mp 1))
\Lambda(\pm y-\alpha+\beta_2+\beta_5/2,\pm m-\theta(\mp 1))
\over \Lambda(2y,2m+1)\Lambda(-2y,-
2m-1)}\\ \nonumber
&&\times\Lambda(\pm y+Q-\alpha-\beta_2+\beta_5/2, \pm m-\theta(\mp 1))
\Lambda(\pm y+\alpha-\beta_2+\beta_5/2,\pm m+1+\theta(\pm 1))\\ \nonumber
&&e^{-i\pi y^2}
{dy\over 2i}\, .
\eea
Collecting all we derive
\bea\label{srul}
&&S_{\alpha\beta_2}(\tilde{\beta}_5)=-S_{0\beta_2}{e^{-{i\pi\over 2}(\Delta_{\beta_5}+1/2)}\over S_0(\beta_5)}\quad\\ \nonumber
&&\times\int_{-i\infty}^{i\infty}\sum_{m=0}^{1}\Lambda(y-Q/2+\beta_2+\beta_5/2, m-1)\Lambda(y+Q/2-\beta_2+\beta_5/2, m+1)\quad\\ \nonumber
&&\times\Lambda(-y-Q/2+\beta_2+\beta_5/2,-m)\Lambda(-y+Q/2-\beta_2+\beta_5/2, -m)e^{2i\pi y(-Q/2+\alpha)} e^{i\pi m} {dy\over i}\, ,
\eea

or using \eqref{l1s2}:

\bea\label{srulik}
&&S_{\alpha\beta_2}(\tilde{\beta}_5)=S_{0\beta_2}{e^{-{i\pi\over 2}\Delta_{\beta_5}}\over S_{\rm R}(\beta_5)}\quad\\ \nonumber
&&\times\int_{-i\infty}^{i\infty}\Big[S_{\rm NS}(y-Q/2+\beta_2+\beta_5/2)S_{\rm NS}(y+Q/2-\beta_2+\beta_5/2)\quad\\ \nonumber
&&\times S_{\rm R}(-y-Q/2+\beta_2+\beta_5/2)S_{\rm R}(-y+Q/2-\beta_2+\beta_5/2)\\ \nonumber
&&+S_{\rm R}(y-Q/2+\beta_2+\beta_5/2)S_{\rm R}(y+Q/2-\beta_2+\beta_5/2)\quad\\ \nonumber
&&\times S_{\rm NS}(-y-Q/2+\beta_2+\beta_5/2)S_{\rm NS}(-y+Q/2-\beta_2+\beta_5/2)\big]
e^{2i\pi y(-Q/2+\alpha)}  {dy\over i}\, .
\eea
Note the following properties of this expression. First we obtain easily
\be\label{sabo11}
S_{\alpha\beta_2}(\tilde{\beta_5})=S_{Q-\alpha,\beta_2}(\tilde{\beta_5})=S_{\alpha, Q-\beta_2}(\tilde{\beta_5})\, .
\ee
Using formula \eqref{symref} in appendix E one can establish
\bea\label{sabo111}
&&S_{\alpha\beta_2}(\widetilde{Q-\beta_5})=S_{\rm R}(-Q+2\alpha+\beta_5)S_{\rm R}(Q-2\alpha+\beta_5)\\ \nonumber
&&\times S_{\rm R}(2Q-\beta_5-2\beta_2)S_{\rm R}(-\beta_5+2\beta_2)
S_{\alpha\beta_2}(\tilde{\beta}_5)\, .
\eea
Combining \eqref{sabo11} and \eqref{sabo111} with definition of the normalization factors \eqref{dozz2} and using formulae \eqref{lsns} in apendix A we obtain that
\be
S^{SL}_{\alpha\beta_2}(\tilde{\beta}_5)={{\cal N}_{R}(\alpha,\beta_5,\alpha)\over {\cal N}_{R}(\beta_2,\beta_5,\beta_2)}S_{\alpha\beta_2}(\tilde{\beta}_5)
\ee
 is invariant under the reflection of all three variables.
\smallskip

\section{Conclusion}
In this paper we calculated the matrix $S_{\alpha\beta_2}(\beta_5)$ in two cases.  These results allow us to conjecture
that in all cases the integral part of the matrix $S_{\alpha\beta_2}(\beta_5)$ will be given by the following function with the various choices of discrete variables $k_a, t_a$, $a=1,2$ and $N$:
\bea\label{funki}
&&F_{\alpha,\beta_2}(\beta_5)(\underline{k},\underline{t},N)=
\int_{-i\infty}^{i\infty}\sum_{m=0}^{1}\Lambda(y-Q/2+\beta_2+\beta_5/2, k_1+m)\\ \nonumber
&&\times\Lambda(y+Q/2-\beta_2+\beta_5/2, k_2+m)\Lambda(-y-Q/2+\beta_2+\beta_5/2,t_1-m)\\ \nonumber
&&\times\Lambda(-y+Q/2-\beta_2+\beta_5/2, t_2-m)e^{2i\pi y(-Q/2+\alpha)} e^{i\pi mN} dy\, .
\eea
Applying some limiting procedures to the finite-difference equation derived in \cite{Apresyan:2022erh}
we obtained that for the parameters $t_1,t_2,k_1,k_2$ obeying the condition
\be
k_1-k_2=t_1-t_2+2l\, ,
\ee
the function \eqref{funki} satisfies the eigenvalue equation:

\bea\label{hamo3}
&&e^{i\pi (L-l)\over 2}{\sin{\pi b\over 2} (2\alpha-Q-\beta_5-(N-M)/b)\over \sin {b\pi\over 2}(2\alpha-Q-(N+l)/b)}F_{\alpha-b/2,\beta_2}(\beta_5)(\underline{k},\underline{t},N+1)\\ \nonumber
&&+e^{-{i\pi (L+l)\over 2}}{\sin{\pi b\over 2}(2\alpha-Q+\beta_5-(N+M)/b)\over \sin {b\pi\over 2}(2\alpha-Q-(N+l)/b)}
F_{\alpha+b/2,\beta_2}(\beta_5)(\underline{k},\underline{t},N-1)\\ \nonumber
&&=2\cos{\pi b\over 2}(Q-2\beta_2-(t_2-t_1)/b)F_{\alpha,\beta_2}(\beta_5)(\underline{k},\underline{t},N)\, ,
\eea
where
\be
M={1\over 2}\sum_{1,2}(t_a+k_a)\, ,\quad L={1\over 2}\sum_{1,2}(t_a-k_a)\, .
\ee
As in the case of Liouville theory one might hope that
the correspondence of the elements of the $S$-move matrix in $N=1$ super Liouville theory with the eigenvalues of the l.h.s. operator could help us
to prove the Moore-Seiberg condition of orthogonality \eqref{msorto}. It is intriguing to study the relation of equation \eqref{hamo3} with the supersymmetric Ruijsenaars-Sutherland problem  \cite{Blondeau-Fournier:2014lba}.


There is also another application of S-move matrix in the AGT correspondence.
Remember that logarithm of one-point toric block in semi-classical limit yields Seiberg-Witten prepotential of  ${\cal N}=2^*$ theory
\cite{Fateev:2009aw,Galakhov:2012gw,Marshakov:2010fx}. Therefore asymptotic form of the S-move matrix of Liouville field theory implements electromagnetic duality of SW prepotential. For Liouville field theory this was studied in \cite{Galakhov:2012gw,Marshakov:2010fx,Kashani-Poor:2014mua}. We are planning to pursue these issues
for S-move matrix in $N=1$ SLFT using relation
between the Nekrasov partition function of  ${\cal N}=2^*$ theory on $S^4/\mathbb{Z}_2$ and $N=1$ SLFT one-point toric blocks found in \cite{Bonelli:2011jx,Bonelli:2011kv}.

We expect that the S-move matrix of super Liouville theory also should play important role in quantization of the super Teichmuller space
\cite{Aghaei:2015bqi,Aghaei:2020otq} and building  supersymmetric analogue of the Virasoro topological field theory studied in
\cite{Collier:2023fwi,Collier:2024mgv}.
\vspace{1cm}

{\bf Acknowledgements.}
The work of Elena Apresyan was supported by  Armenian SCS grants 21AG‐1C024 and 20TTAT-QTa009. We would like also to thank
V. P. Spiridonov for many useful discussions.

\appendix

\section{Properties of supersymmetric hyperbolic and double gamma functions}
The function $S_b(y)$ has the integral representation
\be
S_b(y)=\exp\left(-\int_0^{\infty}\left({\sinh(2y-b-b^{-1})x\over 2\sinh(bx)
\sinh(b^{-1}x)}-{2y-b-b^{-1}\over 2x}\right)\right){dx\over x}\, ,
\ee
and obeys the equations:
\be\label{hp}
{S_b(y+b)\over S_b(y)}=2\sin\pi by\,  ,\quad
{S_b(y+b^{-1})\over S_b(y)}=2\sin{\pi y\over b}.
\ee
Let us mention for further use the following of properties of the
the function $\Lambda(y,m)$. The function $\Lambda(y,m)$ has the
asymptotics \cite{Nieri:2015yia,Sarkissian:2018ppc}:
\be\label{asy1}
\stackreb{\lim}{y\to \infty}\Lambda(y,m)=e^{\mp{\textup{i}\pi\over 2}\left({1\over r}B_{2,2}(y,b,b^{-1})+{m^2\over r}-m+{r\over 6}-{1\over 6r}\right)},
\quad {\rm for}\quad \mathfrak{Im}(y)\to \pm\infty\, ,
\ee
where $B_{2,2}(y,b,b^{-1})$  is the second order Bernoulli polynomial
\be\label{asy2}
B_{2,2}(y;b,b^{-1})=
\left(y-\frac{Q}{2}\right)^2-\frac{b^2+b^{-2}}{12}\, .
\ee
The poles and zeros of the function $\Lambda(y,m)$ are given by the relations \cite{Nieri:2015yia,Sarkissian:2019bmt}:
\be
u^{(m)}_{\rm poles}\in \{ -n_1b -n_2b^{-1}\},\; u^{(m)}_{\rm zeros}\in \{ (n_1+1)b +(n_2+1)b^{-1}\},
\ee
where
$n_1-n_2=m\ {\rm mod }\ r,
\; n_1,n_2 \in {\mathbb Z}_{\geq0}$.

 The function $\Lambda(y,m)$ satisfies the relation:
\be\label{kaz}
{\Lambda(y+b,m-1)\over \Lambda(y,m)}=-2\sin{\pi b(y-mb^{-1})\over r}, \;
{\Lambda(y+b^{-1},m+1)\over \Lambda(y,m)}=2\sin{\pi (y+mb)\over rb}.
\ee
Specializing these properties to the case $r=2$ we derive, that poles and zeros of $S_{NS}(x)$ are given by the set
\be
u^{(0)}_{\rm poles}\in \{ -n_1b -n_2b^{-1}\},\; u^{(0)}_{\rm zeros}\in \{ (n_1+1)b +(n_2+1)b^{-1}\},
\ee
where $n_1-n_2=0\ {\rm mod }\ 2,
\; n_1,n_2 \in {\mathbb Z}_{\geq0}$\, ,
and  poles and zeros of $S_{R}(x)$ are given by the set
\be
u^{(1)}_{\rm poles}\in \{ -n_1b -n_2b^{-1}\},\; u^{(1)}_{\rm zeros}\in \{ (n_1+1)b +(n_2+1)b^{-1}\},
\ee
where $n_1-n_2=1\ {\rm mod }\ 2,
\; n_1,n_2 \in {\mathbb Z}_{\geq0}$.

The relations \eqref{kaz} for the case $r=2$ take the form:
\be\label{kazo}
{S_{NS}(y+b)\over S_{R}(y)}=-2\sin{\pi b(y-b^{-1})\over 2}\,  ,\quad
{S_{NS}(y+b^{-1})\over S_{R}(y)}=2\sin{\pi (y+b)\over 2b}
\ee
and
\be\label{kazol}
{S_{R}(y+b)\over S_{NS}(y)}=2\sin{\pi by\over 2}\,  ,\quad
{S_{R}(y+b^{-1})\over S_{NS}(y)}=2\sin{\pi y\over 2b}.
\ee

We need also relations of sypersymmetric hyperbolic gamma functions with the supersymmetric double gamma functions
\cite{Hadasz:2013bwa,Poghosyan:2016kvd}.

Start with defintion of  $\Gamma_b(x)$ function.

The function $\Gamma_b(x)$ is a close relative of the double Gamma function studied in \cite{gam1,gam2}.
It can be defined by means of the integral representation
\be
\log \Gamma_b(x)=\int_0^{\infty}{dt\over t}\left({e^{-xt}-e^{-Qt/2}\over (1-e^{-bt})(1-e^{-t/b})}-
{(Q-2x)^2\over 8e^t}-{Q-2x\over t}\right)\, .
\ee

Important properties of $\Gamma_b(x)$ are
\begin{enumerate}
\item
Functional equation: $\Gamma_b(x+b)=\sqrt{2\pi}b^{bx-{1\over 2}}\Gamma^{-1}(bx)\Gamma_b(x)$.
\item
Analyticity:  $\Gamma_b(x)$ is meromorphic, poles: $x=-nb-mb^{-1}, n,m\in \mathbb{Z}^{\geq 0}$.
\item
Self-duality:  $\Gamma_b(x)= \Gamma_{1/b}(x)$.
\end{enumerate}

The function $S_b(x)$ may be defined in terms of $\Gamma_b(x)$ as follows
\be\label{sgg}
S_b(x)={\Gamma_b(x)\over  \Gamma_b(Q-x)}\, .
\ee

Similar role in the super Liouville theory play the functions:
\be
\Gamma_{1}(x)\equiv\Gamma_{\rm NS}(x)=\Gamma_b\left({x\over 2}\right)\Gamma_b\left({x+Q\over 2}\right)\, ,
\ee
\be
\Gamma_{0}(x)\equiv\Gamma_{\rm R}(x)=\Gamma_b\left({x+b\over 2}\right)\Gamma_b\left({x+b^{-1}\over 2}\right)\, .
\ee
The sypersymmetric hyperbolic gamma functions are related to them by the formulae similar to \eqref{sgg}
\be\label{lsns}
S_{1}(x)\equiv S_{\rm NS}(x)={\Gamma_{\rm NS}(x) \over \Gamma_{\rm NS}(Q-x)}\, ,\quad
S_{0}(x)\equiv S_{\rm R}(x)={\Gamma_{\rm R}(x) \over \Gamma_{\rm R}(Q-x)}\, .
\ee

\section{Parafermionic integral identities I}
In paper \cite{Sarkissian:2018ppc} the rarefied hyperbolic beta integral has been evaluated\footnote{We put periods $\omega_1$ and $\omega_2$ there equal $b^{-1}$ and $b$ correspondingly and also $\epsilon=0$.}:

\bea\label{betin}
&&\int_{-i\infty}^{i\infty}\sum_{m=0}^{r-1}{\prod_{a=1}^6\Lambda(s_a\pm y,n_a\pm m)
\over \Lambda(\pm 2y, \pm 2m)}{dy\over i}
= 2r\prod_{1\leq a < b \leq 6}
\Lambda(s_a+s_b,n_a+n_b)\, ,\qquad
\eea
where
\be\label{balo}
\sum_{a=1}^6n_a=0\, ,\quad \sum_{a=1}^6 s_a=Q\, .
\ee
Here we used the shorthand notation $\Lambda(y\pm x ,m\pm n)\equiv\Lambda(y+ x ,m+ n)\Lambda(y- x ,m- n)$.

It is shown in \cite{Sarkissian:2018ppc}, that in certain limit \eqref{betin} yields

\bea\label{pf2}
&&\int_{-i\infty}^{i\infty}\sum_{m=0}^{r-1}(-1)^m\prod_{a=1}^3\Lambda(y+f_a,n_a+m)
\Lambda(-y+g_a,l_a-m)
{dy\over i}\quad\quad\quad
\\ \nonumber
&&=(-1)^{n_1+n_2+n_3}r\prod_{a,b=1}^3
\Lambda(f_a+g_b,n_a+l_b)\, ,
\eea
where
\be\label{bal}
\sum_{a=1}^3(n_a+l_a)=0\, ,\quad \sum_{a=1}^3 (f_a+g_a)=Q\, .
\ee
In certain limit \eqref{pf2} leads to another useful identity:
\bea\label{tigr}  \makebox[-2em]{}
&&\int_{-i\infty}^{i\infty}\sum_{m=0}^{r-1}(-)^m e^{2i\pi Nm\over r}
e^{ {2i\pi\over r} yt}
\Lambda(f\pm y,n\pm m)
{dy\over i}
\\ &&  \makebox[2em]{}
=(-1)^{N-n}r
\Lambda(2f,2n)\Lambda(Q/2-f\pm t,-n\pm N)
.\nonumber
\eea
We can show that taking the following limit in \eqref{betin}:
\be
s_5\to \textup{i}\infty,\qquad s_6=Q-s_5-\sum_{j=1}^4s_j\to -\textup{i}\infty
\ee
and using the asymptotics \eqref{asy1} one derives
\bea\label{newidzz}
&&\int_{-i\infty}^{i\infty}\sum_{m=0}^{r-1}{\prod_{a=1}^4\Lambda(s_a\pm y,n_a\pm m)
\over \Lambda(\pm 2y,\pm 2m)}{dy\over i}
= 2r{\prod_{1\leq a < b \leq 4}
\Lambda(s_a+s_b,n_a+n_b)\over \Lambda(\sum_{i=1}^4s_i,\sum_{i=1}^4n_i)}\, .\qquad
\eea
Taking in \eqref{newidzz} further limit
\be
s_4\to \textup{i}\infty
\ee

and again using the asymptotics \eqref{asy1} one gets
\bea\label{betnews}
&&\int_{-i\infty}^{i\infty}\sum_{m=0}^{r-1}{\prod_{a=1}^3\Lambda(s_a\pm y,n_a\pm m)
\over \Lambda(\pm 2y,\pm 2m)}e^{-{i\pi\over r} y^2}e^{-{i\pi\over r} m^2}{dy\over i}
= 2r e^{{i\pi\over r}(s_1s_2+s_2s_3+s_1s_3)}\qquad\quad\\ \nonumber
&&\times e^{{i\pi\over r}(n_1n_2+n_2n_3+n_1n_3)}\Lambda(s_1+s_2,n_1+n_2)\Lambda(s_2+s_3,n_2+n_3)\Lambda(s_1+s_3,n_1+n_3).
\eea
And finally taking the limit
\be
s_3\to \textup{i}\infty
\ee
in \eqref{betnews} and again using the asymptotics \eqref{asy1} we obtain
\bea\label{pf1}
&&\int_{-i\infty}^{i\infty}\sum_{m=0}^{r-1}{\prod_{a=1}^2\Lambda(s_a\pm y,n_a\pm m)
\over \Lambda(\pm 2y,\pm 2m)}e^{-{2i\pi\over r} y^2}e^{-{2i\pi\over r} m^2}{dy\over i}\quad\quad
\\  \nonumber
&&= 2r e^{-{i\pi\over 2r}(s_1-s_2)^2}e^{{i\pi\over 2r}(s_1+s_2)Q}e^{-{i\pi\over 2r}(n_1-n_2)^2}e^{{i\pi\over 2}(n_1+n_2)}\Lambda(s_1+s_2,n_1+n_2).
\eea

\section{Parafermionic integral identities II}

In paper \cite{Apresyan:2022erh} some parafermionic hypergeometric functions were defined.
The first is the parafermionic generalization of the Ruijsenaars function \cite{ruijj}:
\bea\label{jparker}
&&J(\underline{\beta},\underline{l};\underline{\gamma},\underline{t})=\int_{-i\infty}^{i\infty}\sum_{m=0}^{r-1}\prod_{a=1}^4
\Lambda(y+\gamma_a,t_a+m)
\Lambda(-y+\beta_a,l_a-m){dy\over ir}
\ .\qquad
\eea
Here $t_a\, , l_a\in \mathbb{Z}$. Parameters $\gamma_j$, $\beta_j$ and $l_j$, $t_j$ satisfy the constraints:
\begin{equation}\label{mmm}
\sum_{j=1}^4 (\gamma_j+\beta_j)=2Q,\quad \sum_{j=1}^4 (l_j+t_j)=0.
\end{equation}

The second function is :
\bea\label{efuk}
E_{\delta}(s_a,n_a)=\int_{-i\infty}^{i\infty}\sum_{m=0}^{r-1}{\prod_{a=1}^6\Lambda(y+s_a,n_a+m+\delta)
\Lambda(-y+s_a,n_a-m)
\over \Lambda(2y,2m+\delta)\Lambda(-2y,-
(2m+\delta))}{dy\over 2ir}\, .\quad
\eea
We proved in \cite{Apresyan:2022erh} the following relation between \eqref{jparker} and \eqref{efuk}:
\bea\label{paradva}
&&J(\underline{\beta}, \underline{l};\underline{\gamma},\underline{t})=e^{i\pi(t_4+l_4)}\prod_{i=1}^3\Lambda(\beta_i+\gamma_4,l_i+t_4)
\Lambda(\gamma_i+\beta_4,t_i+l_4)\\ \nonumber
&&\times E_{\delta}(\beta_1+\xi,l_1-L_1;\beta_2+\xi,l_2-L_1;\beta_3+\xi,l_3-L_1;\nonumber \\
&&\gamma_1-\xi,t_1-L_2;\gamma_2-\xi,t_2-L_2;\gamma_3-\xi,t_3-L_2)\, ,\quad
\eea
where
\be
L_1={1\over 2}(l_1+l_2+l_3+t_4+\delta)\, ,
\ee
\be
L_2={1\over 2}(t_1+t_2+t_3+l_4+\delta)\, ,
\ee
\be
\xi={1\over 2}(Q-\gamma_4-\sum_{j=1}^3 \beta_j).
\ee
The parameter $\delta=0,1$ should be determined from the requirement that $L_1$ and $L_2$ are integer numbers.
Let us now paparemeterize parameters $\underline{\beta}$, $\underline{\gamma}$ in \eqref{paradva}
in the following way:
\be\label{pr11}
\begin{aligned}
&\gamma_1=A-\alpha_1+Q/2\, ,\\
&\gamma_2=\beta_2=\alpha_2/2-\alpha_3+Q/4\, ,\\
&\gamma_3=\beta_3=\alpha_2/2+\alpha_3+Q/4\, ,
\end{aligned}\qquad
\begin{aligned}
&\gamma_4=-A-\alpha_2+\alpha_1\, ,\\
&\beta_1=A+\alpha_1+Q/2\, ,\\
&\beta_4=-A-\alpha_2-\alpha_1\, .
\end{aligned}
\ee
Obviously parametrized in this way $\underline{\beta}$ and  $\underline{\gamma}$  satisfy \eqref{mmm}.

Now putting (\ref{pr11}) in  (\ref{paradva}) and taking the limit $A\to \infty$, denoting $l_1+t_4=T$, and at the end
renaming indices $2,3\to 1,2$ for all variables, what we can do since variables with indices $1,4$ disappear,  we obtain the identity
\bea\label{pf4}
&&\int_{-i\infty}^{i\infty}\sum_{m=0}^{r-1}\prod_{a=1}^2\Lambda(y+\gamma_a,t_a+m)
\Lambda(-y+\gamma_a,l_a-m)e^{{4i\over r}\pi y\alpha_1} e^{{i\pi m\over r}(2T+\sum_{a=1}^2(t_a+l_a))} dy
=\\ \nonumber
&&\Lambda(2\alpha_1-\gamma_1-\gamma_2+Q,T)
\Lambda(-2\alpha_1-\gamma_1-\gamma_2+Q,-T-\sum_{a=1}^2(t_a+l_a))e^{-i\pi\left[\sum_{a=1}^2(t_a+l_a)+{2\over r}\left(\gamma_1\gamma_2-\alpha_1^2\right)\right]}
\\ \nonumber
&&\times
\int_{-i\infty}^{i\infty}\sum_{m=0}^{r-1}{\prod_{a=1}^2\Lambda(y+\gamma_a-\alpha_1,l_a-M_1+m+\delta)
\Lambda(-y+\gamma_a-\alpha_1,l_a-M_1-m)
\over \Lambda(2y,2m+\delta)\Lambda(-2y,-
(2m+\delta))}\\ \nonumber
&&\times\prod_{a=1}^2\Lambda(y+\gamma_a+\alpha_1,t_a-M_2+m+\delta)
\Lambda(-y+\gamma_a+\alpha_1,t_a-M_2-m)e^{-{2i\pi\over r} y^2}e^{-{2i\pi \over r}(m^2+m\delta)}
{dy\over 2}\\ \nonumber
&&\times e^{{i\pi \over 2r}\left[l_1^2+l_2^2+t_1^2+t_2^2-2M_1^2-2M_2^2+2(T+l_1+l_2)^2-(t_1+t_2)^2-(l_1+l_2)^2\right]}\, ,
\eea
where
\be
M_1={1\over 2}(T+l_1+l_2+\delta)\, ,
\ee
\be
M_2=\delta-M_1\, ,
\ee
and $\delta=0,1$ is determined from the requirement that $M_1$ and therefore also $M_2$ are integer.
This formula for $r=1$ case can be found in \cite{Spiridonov:2011hf,Vartanov:2013ima}.

\section{Parafermionic integral identities III}

Now let us set in (\ref{paradva})
\be
\beta_4=-{\lambda\over 3}-2A+Q/2\,  ,\quad
\gamma_1=-{\lambda\over 3}+A+Q/2\, ,\quad
\gamma_2=-{\lambda\over 3}+A+Q/2\, .
\ee
Taking the limit $A\to \infty$, we obtain, after  renaming at the end the variables $\gamma_3,\gamma_4, t_3,t_4\to\gamma_1,\gamma_2, t_1,t_2$,
the identity
\bea\label{pf5}
&&\int_{-i\infty}^{i\infty}\sum_{m=0}^{r-1}\prod_{a=1}^2\Lambda(y+\gamma_a,t_a+m)
\prod_{a=1}^3\Lambda(-y+\beta_a,l_a-m)e^{{i\pi \over r}(\lambda y-{y^2\over 2})}
e^{{i\pi  \over r}\left[m(t_1+t_2+\sum_{i=1}^3l_i+{3r\over 2})-{m^2\over 2}\right]}
dy\quad\quad\quad\\ \nonumber
&&= \int_{-i\infty}^{i\infty}\sum_{m=0}^{r-1}{\prod_{a=1}^3\Lambda(y+\beta_a+\eta,l_a-L_1+m+\delta)
\Lambda(-y+\beta_a+\eta,l_a-L_1-m)
\over \Lambda(2y,2m+\delta)\Lambda(-2y,-
(2m+\delta))}\\ \nonumber
&&\times\Lambda(y+\gamma_1-\eta,t_1-L_2+m+\delta)
\Lambda(-y+\gamma_1-\eta,t_1-L_2-m)e^{-{2i\pi\over r} y^2}
e^{-{2i\pi\over r} (m^2+m\delta)}
{dy\over 2}\\ \nonumber
&&\times e^{{i\pi \over 2r}(\lambda^2+Q^2/4-Q(\lambda+\gamma_1))}e^{{i\pi\over 2r}\left[4L_2^2-4L_2(t_1+\delta)+2\delta t_1+t_1^2\right]}e^{{i\pi\over 2}\left[-t_1+2(l_1+l_2+l_3)\right]}\prod_{a=1}^3\Lambda(\beta_a+\gamma_2;l_a+t_2)\, ,
\eea
where
\be\label{ball}
\sum_{i=1}^2\gamma_i+\sum_{i=1}^3\beta_i=Q/2+\lambda\, ,
\ee
and
\be
2\eta=Q-\gamma_2-\sum_{i=1}^3\beta_i=Q/2-\lambda+\gamma_1\, ,
\ee

\be
L_1={1\over 2}(l_1+l_2+l_3+t_2+\delta)\, ,
\ee
\be
L_2=\delta-L_1\, ,
\ee

and $\delta=0,1$ is determined from the requirement that $L_1$ and therefore also $L_2$ are integer.
For $r=1$ case this formula can be found in \cite{Vartanov:2013ima}.

\section{Parafermionic integral identities IV}

We proved also in \cite{Apresyan:2022erh} that \eqref{jparker} satisfies the transformation property

\bea\label{jjsym11}
&&J(\underline{\beta},\underline{l};\underline{\gamma},\underline{t})=e^{\textup{i}\pi
\left[t_3+t_4+l_1+l_2\right]}J(\underline{\tilde{\beta}},\underline{\tilde{l}};\underline{\tilde{\gamma}},\underline{\tilde{t}})\times\\ \nonumber
&&\prod_{j,k=1}^2\Lambda(\gamma_j+\beta_k,t_j+l_k)
\prod_{j,k=3}^4\Lambda(\gamma_j+\beta_k,t_j+l_k)\, ,
\eea
\bea
&&\tilde{\beta}_a=\beta_a+\Theta(a)\eta\, , \quad \tilde{l}_a=l_a-(\Theta(a)+1)K\, , \quad a=1,2,3,4\, ,\\ \nonumber
&&\tilde{\gamma}_a=\gamma_a+\Theta(a)\eta\, , \;\quad \tilde{t}_a=t_a-(\Theta(a)-1)K\, ,\quad a=1,2,3,4\, ,
\eea
where $\Theta(a)$  is the sign function taking the values
\be\label{thesig}
\Theta(a)=1\, , \quad a=1,2 \quad {\rm and} \quad \Theta(a)=-1\, , \quad a=3,4\, ,
\ee

and
\be
K={1\over 2}(t_1+t_2+l_1+l_2)=-{1\over 2}(t_3+t_4+l_3+l_4)\, ,
\ee
\be
\eta={1\over 2}(Q-\gamma_1-\gamma_2-\beta_1-\beta_2)\, .
\ee
Setting in \eqref{jjsym11}:
\bea\label{pr113}
&&\gamma_1=A-\alpha_1+Q/2\, ,\quad\beta_1=A+\alpha_1+Q/2\, ,\\ \nonumber
&&\gamma_2=-A-\alpha_2+\alpha_1+Q/2\, ,\quad\beta_2=-A-\alpha_1-\alpha_2+Q/2\, ,\\ \nonumber
&&\gamma_3=\beta_3=\alpha_2/2-\alpha_3\, ,\quad\gamma_4=\beta_4=\alpha_2/2+\alpha_3\, ,
\eea

and denoting
\be
N={1\over 2}(t_2+l_1-t_1-l_2)\, ,
\ee
we obtain in the limit $A\to \infty$
\bea\label{symref}
&&\sum_{m\in \mathbb{Z}_r}\int_{-i\infty}^{i\infty}\Lambda( y+\alpha_2/2+\alpha_3,t_4+m)\Lambda(-y+\alpha_2/2+\alpha_3,l_4-m)\\ \nonumber
&&\times\Lambda(y+\alpha_2/2-\alpha_3,t_3+m)\Lambda(- y+\alpha_2/2-\alpha_3,l_3-m)e^{{4i\pi\over r} y\alpha_1}e^{{2miN\pi\over r}}  {dy\over i}=\\ \nonumber
&& e^{{2iNK\pi\over r}}e^{i\pi(t_3+t_4-N)}\Lambda(Q-2\alpha_1-\alpha_2,K-N)\Lambda(Q+2\alpha_1-\alpha_2,K+N)
\\ \nonumber
&&\times\Lambda(\alpha_2-2\alpha_3,t_3+l_3)\Lambda(\alpha_2+2\alpha_3,t_4+l_4)\Lambda(\alpha_2,t_3+l_4)\Lambda(\alpha_2,t_4+l_3)\\ \nonumber
&&\times\sum_{m\in \mathbb{Z}_r}\int_{-i\infty}^{i\infty}\Lambda(y+(Q-\alpha_2)/2+\alpha_3,t_4+2K+m)\Lambda(- y+(Q-\alpha_2)/2+\alpha_3,l_4-m)\\ \nonumber
&&\times\Lambda(y+(Q-\alpha_2)/2-\alpha_3,t_3+2K+m)\Lambda(- y+(Q-\alpha_2)/2-\alpha_3,l_3-m)e^{{4i\pi\over r} y\alpha_1}e^{{2miN\pi\over r}}  {dy\over i}\, ,
\eea
where $K+N\in \mathbb{Z}$ and $K-N\in \mathbb{Z}$\, .

\end{document}